\documentclass[paper]{JHEP3}
\usepackage{epsfig}
\usepackage{amsmath}
\usepackage{amssymb}
\usepackage{amsbsy}
\usepackage{enumerate}
\usepackage{url}
\usepackage{xspace}
\usepackage{footmisc}
\newlength{\wfig}
\newlength{\hfig}
\newlength{\hfigs}
\newcommand{\Wbb}{{\tt Wbb}}
\newcommand{\Wbbj}{{\tt Wbbj}}

\newcommand\pTWbb{p_{\rm\scriptscriptstyle T}^{\rm \scriptscriptstyle Wbb}}
\newcommand\yWbb{y^{\scriptscriptstyle \rm Wbb}}
\newcommand\mb{m_{\scriptscriptstyle \rm b}}

\def\beq{\begin{equation}}
\def\beqn{\begin{eqnarray}}
\def\eeq{\end{equation}}
\def\eeqn{\end{eqnarray}}

\newcommand\HERWIG{{\tt HERWIG}}

\newcommand\HWpp{{\tt HERWIG++}}
\newcommand\Herwigpp\HWpp
\newcommand\HERWIGPP\HWpp
\newcommand\PYTHIA{{\tt PYTHIA}}
\newcommand\PYTHIASIX{{\tt PYTHIA}\xspace}

\newcommand\KRA{K_{\scriptscriptstyle \rm R}}
\newcommand\KFA{K_{\scriptscriptstyle \rm F}}

\newcommand\muf{\mu_{\sss\rm F}}
\newcommand\mur{\mu_{\sss\rm R}}

\def\lq{\left[} 
\def\rq{\right]}

\def\({\left(} 
\def\){\right)}

\newcommand\sss{\mathchoice%
{\displaystyle}%
{\scriptstyle}%
{\scriptscriptstyle}%
{\scriptscriptstyle}%
}

\newdimen\hbigcirc
\newdimen\wbigcirc

\newdimen\figwidth

\figwidth=\textwidth

\newcommand\ep{\epsilon}

\newcommand\as{\alpha_{\sss\rm S}}

\newcommand\pt{p_{\sss \rm T}}
\newcommand\pT{p_{\sss \rm T}}

\newcommand\kt{k_{\sss\rm T}}

\newcommand\MCatNLO{{\tt MC@NLO}}

\newcommand     \MSB            {\ifmmode {\overline{\rm MS}} \else
                                 $\overline{\rm MS}$\fi}

\newcommand\CA{C_{\sss\rm A}}

\newcommand\TF{T_{\sss\rm F}}

\newcommand\nf{n_{\rm f}}
\newcommand\nlf{n_{\rm lf}}
\newcommand\asnf{\as^{\rm (\nf)}}
\newcommand\asnlf{\as^{\rm (\nlf)}}

\newcommand\POWHEGBOX{{\tt POWHEG BOX}}
\newcommand\GOSAM{{\tt GoSam}}
\newcommand\FASTJET{{\tt Fastjet}}
\newcommand\FORM{{\tt FORM}}

\newcommand\SAMURAI{{\tt SAMURAI}}
\newcommand\GOLEM{{\tt Golem95}}
\newcommand\NINJA{{\tt Ninja}}
\newcommand\SPINNEY{{\tt SPINNEY}}
\newcommand\MG{{\tt MadGraph4}}

\newcommand\MCFM{{\tt MCFM}}
\newcommand\MINLO{{\tt MiNLO}}
\newcommand\MiNLO{{\tt MiNLO}}

\newcommand\QGRAF{{\tt QGRAF}}

\newcommand\ttilde{\raise.17ex\hbox{$\scriptstyle\mathtt{\sim}$}}

\def\ord#1{{\cal O}\(#1\)}

\newcommand{\mathd}{\mathrm{d}}

\newcount\minutes 
\newcount\scratch 
\def\timestamp{%
\scratch=\time 
\divide\scratch by 60 
\edef\hours{\the\scratch} 
\multiply\scratch by 60 
\minutes=\time 
\advance\minutes by -\scratch 
---$\,$\hours:\null 
\ifnum\minutes< 10 0\fi 
\the\minutes}

\preprint{MPP-2015-9}

\title{$\boldsymbol{Wb\bar{b}j}$ production at NLO with  POWHEG+MiNLO
\vspace{-0.5cm}
}

\author{Gionata Luisoni\\ 
  Max-Planck Institut f{\"u}r Physik, F\"ohringer 6, D-80805 Munich, Germany\\
  E-mail: \email{luisonig@mpp.mpg.de}
}

\author{Carlo Oleari\\
  Universit\`a di Milano-Bicocca and INFN, Sezione di Milano-Bicocca\\
  Piazza della Scienza 3, 20126 Milano, Italy\\
  E-mail: \email{carlo.oleari@mib.infn.it}}

\author{Francesco Tramontano\\
  Universit\`a di Napoli ``Federico II'' and INFN, Sezione di Napoli,\\
  Complesso di Monte Sant'Angelo, via Cintia, 80126 Napoli, Italy\\
  E-mail: \email{francesco.tramontano@na.infn.it}
}
\abstract{ We present a next-to-leading order plus parton-shower event
  generator for the production of a $W$ boson plus two bottom quarks
  and a jet at hadron colliders, implemented in the \POWHEGBOX{}
  framework. Bottom-mass effects and spin correlations of the decay
  products of the $W$ boson are fully taken into account.
  The code has been automatically generated using the two available
  interfaces to \MG{} and \GOSAM, the last one updated to a new version.
  We have applied the \MINLO{} prescription to our $Wb\bar{b}j$ calculation,
  obtaining a finite differential cross section also in the limit of
  vanishing jet transverse momentum.
  Furthermore, we have compared several key distributions for $Wb\bar{b}j$
  production with those generated with a next-to-leading order plus
  parton-shower event generator for $Wb\bar{b}$ production, and studied their
  factorization- and renormalization-scale dependence. Finally, we have
  compared our results with recent experimental data from the ATLAS and CMS
  Collaborations.  }


\begin{document}

\section{Introduction}
\label{sec:intro}
The production of a $W$ boson in association with two $b$ jets at hadron
collider has many interesting experimental and theoretical facets. On the
experimental side, more interest for this process has been recently driven by
the discovery of a light scalar
particle~\cite{Aad:2012tfa,Chatrchyan:2012ufa}, whose characteristics point
to make it a suitable candidate for being the Higgs boson responsible for the
spontaneous symmetry breaking of the Standard Model. In this respect,
$Wb\bar{b}+X$ is an irreducible background for $HW$ production, with the
Higgs boson decaying into $b$ quarks. In addition, it is also a background to
single top and top-pair production, where the top quark decays into a $Wb$
pair, and to many new physics searches.

On the theoretical side, the calculation of differential cross sections at
hadronic colliders in the presence of massive quarks is surely more
challenging than with massless partons. In addition, the presence of
logarithmic-enhanced terms of ratios of the quark mass over
higher scales may invalidate the perturbative expansion in the strong
coupling constant $\as$ (see also ref.~\cite{Maltoni:2012pa} for a recent
review of the subject).

$Wb\bar{b}$ production at next-to-leading order~(NLO) in QCD has been studied
for a while~\cite{Ellis:1998fv, FebresCordero:2006sj, Cordero:2009kv,
  Badger:2010mg, Frederix:2011qg}. All these calculations were performed in
the so-called 4-flavour scheme, where the $b$ quark is treated as massive
(except for ref.~\cite{Ellis:1998fv}, where the bottom quark is massless) and
there is no direct contribution from the $b$ parton-distribution function in
the incoming hadrons. $Wb+X$ production in the 5-flavour scheme, where the
$b$ quark is treated as massless and there is a contribution from the $b$
parton-distribution function, is discussed, for example, in
refs.~\cite{Campbell:2006cu, Campbell:2008hh, Caola:2011pz}.  $Wb\bar{b}$
production is also available in NLO+parton-shower~(NLO+PS) event generators
such as the \POWHEGBOX~\cite{Oleari:2011ey} and
\MCatNLO~\cite{Frederix:2011qg}.

In this paper we present a NLO calculation for $Wb\bar{b}$~+~1~jet production
interfaced with the POWHEG method~\cite{Nason:2004rx, Frixione:2007vw} and
distributed as part of the \POWHEGBOX{} package~\cite{Alioli:2010xd}.
Bottom-mass effects and spin correlations of the leptonic decay products of
the $W$ boson have been fully taken into account.  In the following we will
refer to this event generator as \Wbbj.

The Born, real, spin- and colour-correlated Born amplitudes have been
generated automatically using the interface of
\MG~\cite{Stelzer:1994ta,Alwall:2007st} to the
\POWHEGBOX~\cite{Campbell:2012am}. The virtual contribution has also been
computed automatically using the interface~\cite{Luisoni:2013cuh} to
\GOSAM~\cite{Cullen:2011ac,Cullen:2014yla}. 

With the straightforward use of these two interfaces we have also generated a
new code for $Wb\bar{b}$ production at NLO, with exact spin correlations in
the decay of the $W$ boson into leptons. We will refer to this event
generator as \Wbb{}.  In ref.~\cite{Oleari:2011ey}, $Wb\bar{b}$ production
was interfaced with the POWHEG method, and the $W$ decay was simulated in an
approximated way.  We made several comparisons between the generator
described in ref.~\cite{Oleari:2011ey} and the new \Wbb{} one, studying
angular and transverse-momentum distributions of the $W$ decay products, and
found no sizable differences.

The choice of factorization and renormalization scale(s) for $Wb\bar{b}+X$
production is a debated issue in the scientific literature: in fact, it is
well known that NLO corrections to $Wb\bar{b}$ production are quite
large~\cite{Cordero:2009kv}, due to the opening of gluon-initiated channels
at NLO.  A separate paper will be needed to discuss scale dependence more
thoroughly.  In this paper, we apply the \MINLO~\cite{Hamilton:2012np}
procedure to our calculation, and we leave POWHEG and \MINLO{} to choose two
of the scales at which the strong coupling constant is evaluated. This
process, in fact, starts at order $\as^3\, \alpha_{\sss EM}^2$ and gets
contributions up to order $\as^4 \,\alpha_{\sss EM}^2$ at next-to-leading
order.
The advantage of using \MINLO{}  is twofold:
\begin{enumerate}

\item
First of all, we are left only with the choice of the scale(s) for the
primary process, i.e.~the underlying flavour and kinematic configuration
after the clusterization operated by the \MINLO{} procedure\footnote{In the
  \MiNLO{} framework, by primary process we mean the process before any
  branching has occurred, i.e.~$H$ production in $H$ + jets, and $Wb\bar{b}$
  production in the present case.}.

\item
We can provide event samples that are NLO+PS accurate for $Wb\bar{b}j$ production
(this is by construction) but that display, at the same time, a finite
differential cross section in the limit of the accompanying jet $j$ becoming
soft or collinear, overlapping in this way to the \Wbb{} results.
\end{enumerate}

The paper is organized as follows: in sec.~\ref{sec:ingredients} we recall
all the ingredients that are necessary in order to build the NLO+PS code
in the \POWHEGBOX, and some technical detail on the change of the
renormalization scheme. In sec.~\ref{sec:minlo}, we illustrate the
modifications to the original \MINLO{} procedure for the case at hand, and
the scale choice(s) for the primary process. In sec.~\ref{sec:results} we
study some key distributions at the NLO and Les Houches event level, and we
compare the \Wbb{} generator with the \Wbbj+\MINLO{} one. We present a comparison
with CMS and ATLAS data in sec.~\ref{sec:data} and we give our conclusions in
sec.~\ref{ref:conclusions}.

\section{Born, real and virtual contributions}
\label{sec:ingredients}
The code for the computation of the amplitudes for $Wb\bar{b}$ and
$Wb\bar{b}j$ production was generated using the existing interfaces of the
\POWHEGBOX{} to \MG{} and \GOSAM{}~\cite{Cullen:2011ac} presented in
refs.~\cite{Campbell:2012am, Luisoni:2013cuh}. The one-loop amplitudes are
generated with the new version 2.0 of \GOSAM{}~\cite{Cullen:2014yla}, that
uses \QGRAF{}~\cite{Nogueira:1991ex}, \FORM~\cite{Kuipers:2012rf} and
\SPINNEY{}~\cite{Cullen:2010jv} for the generation of the Feynman
diagrams. They are then computed at running time with
\NINJA{}~\cite{vanDeurzen:2013saa,Peraro:2014cba}, which is a reduction
program based on the Laurent expansion of the
integrand~\cite{Mastrolia:2012bu}, and using {\tt
  OneLOop}~\cite{vanHameren:2010cp} for the evaluation of the scalar one-loop
integrals.  
For unstable phase-space points, the reduction automatically
switches to \GOLEM{}~\cite{Cullen:2011kv}, that allows to compute the same
one-loop amplitude evaluating tensor integrals. Alternatively, the
traditional integrand-reduction method~\cite{Ossola:2006us}, extended to
$D$-dimensions~\cite{Ellis:2007br}, as implemented in
\SAMURAI{}~\cite{Mastrolia:2010nb}, can be used.

We point out that a numeric calculation for $Wb\bar{b}j$ was performed in
ref.~\cite{Reina:2011mb} too, with the $W$ boson treated as stable, and in
ref.~\cite{vanDeurzen:2013saa}, with full spin correlations in the decay of
the $W$ boson. In addition, there are other automated codes that can generate
the virtual contributions (see, for example, ref.~\cite{Alwall:2014hca}).

In order to run the code, the user has only to select the sign of the $W$
boson, i.e.~$W^+$ or $W^-$, and its leptonic decay mode, i.e.~electronic,
muonic or tauonic decay, in the \POWHEGBOX{} input file. Other flags to
control the \POWHEGBOX{} behavior are documented in previous implementations
and in the {\tt Docs} directory.

\subsection{The decoupling and \MSB{} schemes}

When performing a fixed-order calculation with massive quarks, one can define
two consistent renormalization schemes that describe the same physics: the
usual \MSB{} scheme, where all flavours are treated on equal footing, and a
mixed scheme~\cite{Collins:1978wz}, that we call decoupling scheme, in which
the $\nlf$ light flavours are subtracted in the \MSB{} scheme, while the
heavy-flavour loop is subtracted at zero momentum. In this scheme, the heavy
flavour decouples at low energies.

The virtual contributions generated by \GOSAM{} are computed in the
decoupling scheme. This means that, since we are dealing with bottom-quark
production, we have the correct virtual contributions if the strong coupling
constant $\as$ is running with 4 light flavours, and if the appropriate
parton distribution functions~(pdfs) do not include the bottom quark in
the evolution.

To make contact with other results expressed in terms of the $\MSB$ strong
coupling constant, running with 5 light flavours, and with pdfs with 5
flavours, we prefer to change our renormalization scheme and to switch to the
$\MSB$ one.

The procedure for such a switch is well known, and was discussed in
ref.~\cite{Cacciari:1998it} (see Appendix~\ref{sec:change_scheme} for a quick
review of this procedure).  For $Wb\bar{b}j$ production, we need to add to
the NLO cross section, computed in the decoupling scheme, the following two
terms:
\begin{itemize}
\item  to the $qq$ initial-state channel
\begin{equation}
- 2\,\TF \,\frac{\as}{2\pi}\, \log \(\frac{\mur^2}{\mb^2}\) \, {\cal B}_{qq}
\end{equation}
\item to the $qg$ initial-state channel
\begin{equation}
2\,\TF \,\frac{\as}{2\pi} \lq \frac{1}{3}\log\(\frac{\muf^2}{\mb^2}\)
-\log\(\frac{\mur^2}{\mb^2}\) \rq {\cal B}_{qg}
\end{equation}
\end{itemize}
where ${\cal B}_{qq}$ and ${\cal B}_{qg}$ are the squared Born amplitude for
the corresponding initial states,  $\mur$ and $\muf$ are the
renormalization and factorization scale, respectively, and $\mb$ is the
bottom-quark mass.

\subsection{LO and NLO comparisons}
In this section we perform a comparison between the leading order~(LO) and
next-to-leading order results for $Wb\bar{b}j$ production, using a static
renormalization and factorization scale.
\begin{figure}[htb]
\begin{center}
\includegraphics[width=0.49\textwidth]{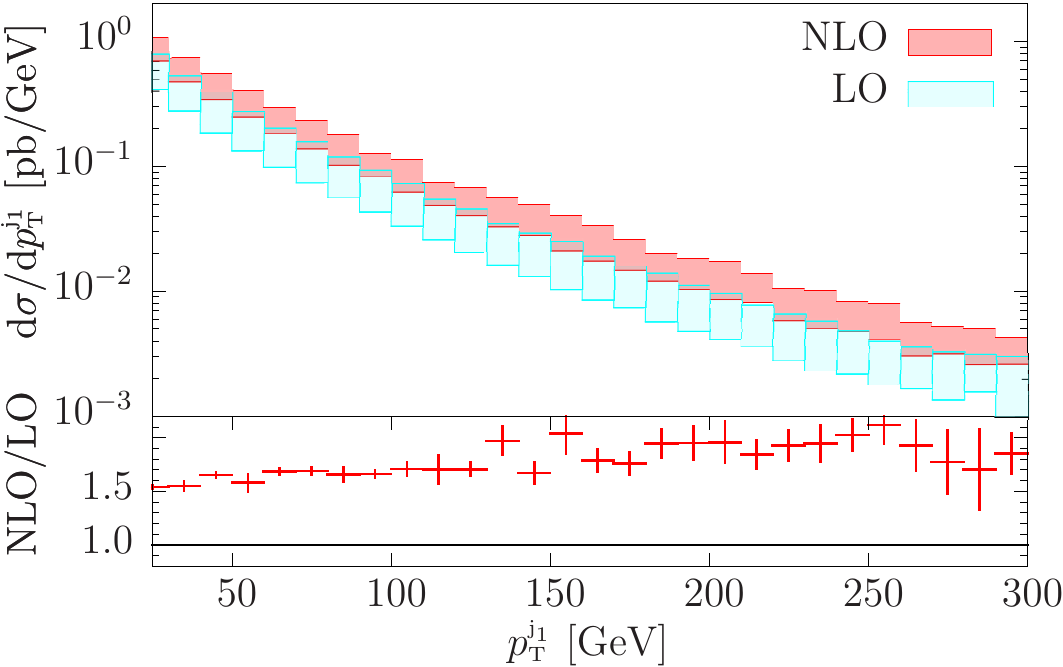}
\includegraphics[width=0.49\textwidth]{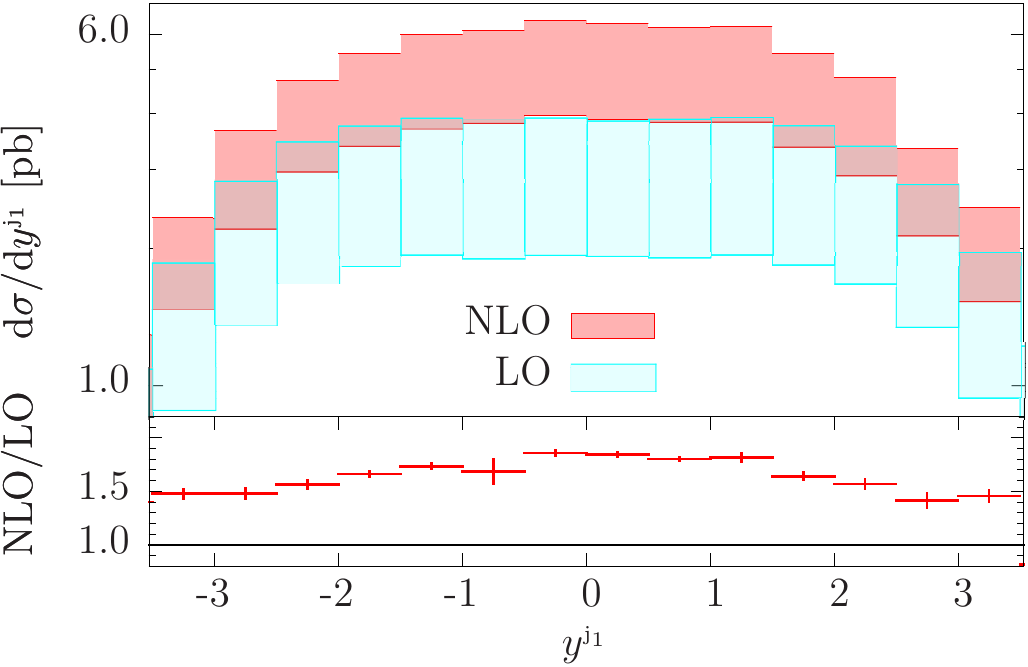}
\end{center}
\caption{Transverse momentum~(left) and rapidity distribution~(right) of the
  hardest jet in $Wb\bar{b}j$ production.  NLO and LO scale-variation bands
  are obtained as explained in the text. In the lower insert, we plot the
  ratio of the NLO over the LO result, computed with the central scale. The
  corresponding integration statistical error is also shown.}
\label{fig:NLO_LO} 
\end{figure}
We present results for $W^+$ production at the LHC at 14~TeV. Similar results
can be obtained for $W^-$ production.  We have set the renormalization and
factorization scale equal to $\mu$
\begin{equation}
\mur=\muf=\mu
\end{equation}
and varied $\mu$ in the range $\mu = \{m_{\sss W}/2, m_{\sss W}, 2\, m_{\sss
  W}\}$. Jets and other physical parameters are defined as in
sec.~\ref{sec:results} and a minimum cut on the hardest jet of 25~GeV has
been imposed, in order to have finite distributions.  The total cross
sections for the central scale are
\beq
\sigma_{\sss\rm LO} = 17.3~{\rm pb}, \qquad \qquad \sigma_{\sss\rm NLO} = 28.4~{\rm pb}.
\end{equation}
By varying the renormalization and factorization scale we have a 46\%
variation for the LO total cross and a 29\% variation at NLO, showing a
significative reduction of the scale dependence.  In fig.~\ref{fig:NLO_LO} we
compare two differential distributions, i.e.~the transverse momentum and
rapidity of the hardest jet, at leading and next-to-leading order. A
reduction on the scale dependence is clear in the two panels.


In the rest of the paper we abandon the use of a fixed scale, since we leave
to \MiNLO{}~\cite{Hamilton:2012np} the choice of scales. Being interested in
a shower Monte Carlo simulation, the most appropriate scale for the
evaluation of the strong coupling constant associated with the emission of a
jet is the transverse momentum of the jet itself, provided the suitable
Sudakov form factor is attached, as done by the \MiNLO{} procedure. We will
further illustrate our use of the \MiNLO{} procedure in the next section.

\section{MiNLO}
\label{sec:minlo}

The \MiNLO{} procedure~\cite{Hamilton:2012np} has already been applied
successfully to several production processes: $H/Z/W
j$~\cite{Hamilton:2012rf}, $W/Zjj$~\cite{Campbell:2013vha},
$HVj$~\cite{Luisoni:2013cuh}, trijet~\cite{Kardos:2014dua},
$W\gamma$~\cite{Barze:2014zba} production.
In $H/Z/W j$ and $HVj$ production, a slightly modified version of the
original \MiNLO{} formalism~\cite{Hamilton:2012rf} allowed to reach NLO
accuracy for inclusive quantities with one less jet too, i.e.~$H/W/Z$ and
$HV$ production, respectively.  With the same modified version and with the
additional knowledge of the fixed NNLO differential cross
section~\cite{Grazzini:2008tf, Catani:2009sm}, Higgs
boson~\cite{Hamilton:2013fea, Hamilton:2015nsa} and
Drell-Yan~\cite{Karlberg:2014qua} production could be simulated at NNLO+PS
accuracy.

Unfortunately, for $Wb\bar{b}j$ production, no such modification of the
\MINLO{} Sudakov form factor is known (due to the presence of coloured
particles in the final state that complicates the structure of the
resummation formulae), and we cannot demonstrate that we can generate an
event sample with NLO accuracy for $Wb\bar{b}$ production too. We will limit
ourselves to show that we can generate a $Wb\bar{b}j$ event sample with
finite differential cross section down to the transverse momentum of the
hardest jet $j$ going to zero, and to show that it agrees fairly well with
the cross section obtained with the \Wbb{} event generator for $Wb\bar{b}$
production.

Our last remark concerns the clusterization procedure operated by \MINLO{}:
since no collinear singularities are associated with the final-state $b$
quarks, the clusterization procedure is not applied to the heavy quarks. In
this way, if the event gets clustered, then we have to deal with the
kinematics of a $Wb\bar{b}$ configuration, otherwise we have a $Wb\bar{b}j$
one.

\subsection{\Wbb{} and \Wbbj{}+\MiNLO{} scale choice}
\label{sec:scale_choice}

$Wb\bar{b}$ and $Wb\bar{b}j$ production are multi-scale processes. The
investigation of the dependence of the differential cross section on
different scale choices is fundamental to assess the uncertainties on the
theoretical predictions and in the comparison with the experimental data.  In
sec.~\ref{sec:results}, we limit ourselves to present results at a given
dynamical scale and we compute scale-variation bands around it.
In sec.~\ref{sec:data}, we present a few results at a different scale when we
compare our theoretical predictions with the ATLAS data.

\noindent
Our default scale choice is the following:
\begin{enumerate}
\item
In the \Wbb{} generator, we have set the renormalization and factorization scales
equal to
\begin{equation}
\label{eq:Wbb_scales}
\mur=\muf= \mu\equiv\frac{E_B}{4}\,, {\rm \quad where \quad } E_B =
\sqrt{\hat{s}}\,, {\rm \quad and \quad } \hat{s} = (p_{\sss\rm W}
+ p_{\sss\rm b}+p_{\sss\rm \bar b})^2,
\end{equation}
where $p_{\sss\rm W}$, $p_{\sss\rm b}$ and $p_{\sss\rm \bar b}$ are the
momenta of the $W$ boson, of the $b$ and $\bar{b}$ quark respectively, at the
underlying-Born level, i.e.~the kinematic configuration on top of which the
\POWHEGBOX{} attaches the hard radiated parton with the appropriate Sudakov
form factor.

\item
In the \Wbbj+\MINLO{} generator, we have the freedom to set the scale(s) of the
primary process. We have fixed this scale to be the same scale $E_B$ of
eq.~(\ref{eq:Wbb_scales}), where now $p_{\sss \rm W}$, $p_{\sss \rm b}$ and
$p_{\sss\rm \bar b}$ are the momenta of the $W$, $b$ and $\bar{b}$ in the
primary process if there has been a clusterization.  If the event has not
been clustered by the \MINLO{} procedure, i.e.~if the underlying Born
$Wb\bar{b}j$ process is not clustered by \MINLO{}, we take as scale the
partonic center-of-mass energy of the event.
\end{enumerate}
With these scale choices, we have good agreement between the the \Wbb{} and
\Wbbj+\MINLO{} results, as will be shown in
sec.~\ref{sec:Wbb_Wbbj_comparisons}\footnote{We leave further investigation
  on the choice of scales to a forthcoming paper.}. The fact that the scale
in the \Wbb{} generator is smaller than the scale of the primary process in
the \Wbbj+\MINLO{} generator is a common feature of the \MINLO{}
procedure. In fact, it has already been observed in vector-boson and Higgs
boson production plus jet~\cite{Hamilton:2012np, Campbell:2013vha,
  Luisoni:2013cuh}, and in trijet production~\cite{Kardos:2014dua}.  

\section{Results}
\label{sec:results}
In this section, we present our findings for the LHC at 7~TeV. We have set
the $b$ quark mass at the value $m_b=4.75$~GeV and we have used the
MSTW2008~\cite{Martin:2009iq} pdf. Clearly, any other pdf could have been
used.

The CKM matrix has been set to
\begin{equation}
\begin{array}{c}
\\
V_{\rm \sss CKM}=
\end{array}
\begin{array}{c c}
& d\quad\quad\ \ s\ \ \quad\quad b \\
\begin{array}{c}
u\\
c\\
t
\end{array} 
&
\left(
\begin{array}{c c c}
0.97428 & 0.2253 & 0.00347\\
0.2252 & 0.97345 & 0.041\\
0.00862 & 0.0403 & 0.999152\end{array}
\right).
\end{array}
\end{equation}
Since the experimental data for $Wb\bar{b}$ production are presented as
summed over $W^+$ and $W^-$ production, we do the same in our plots.

Jets are reconstructed using the anti-$\kt$ algorithm~\cite{Cacciari:2008gp}
as implemented in the \FASTJET{} package~\cite{Cacciari:2005hq,
  Cacciari:2011ma}, with jet radius $R=0.7$.  A minimum transverse-momentum
cut of 1~GeV has been imposed to all the jets. No cuts on $b$-jets have
been imposed.  In all our results, the \MiNLO{} procedure has always been
turned on and events have been showered and hadronized by \PYTHIASIX~(version
6.4.25)~\cite{Sjostrand:2006za}, with the AMBT1 tune (call to {\tt
  PYTUNE(340)}).

\subsection{NLO and Les Houches event comparisons}

In this section, we compare a few interesting kinematic distributions at the
NLO level and at the Les Houches event~(LHE) level, i.e.~after the first hard
emission generated with the POWHEG method.
\begin{figure}[htb]
\begin{center}
\includegraphics[width=0.49\textwidth]{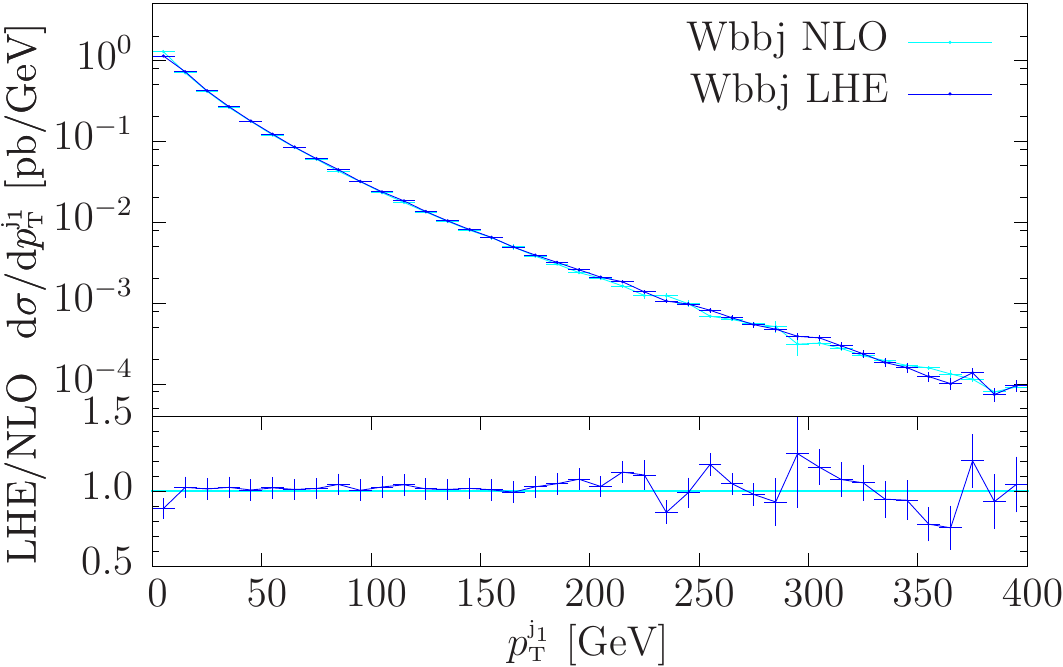}
\includegraphics[width=0.49\textwidth]{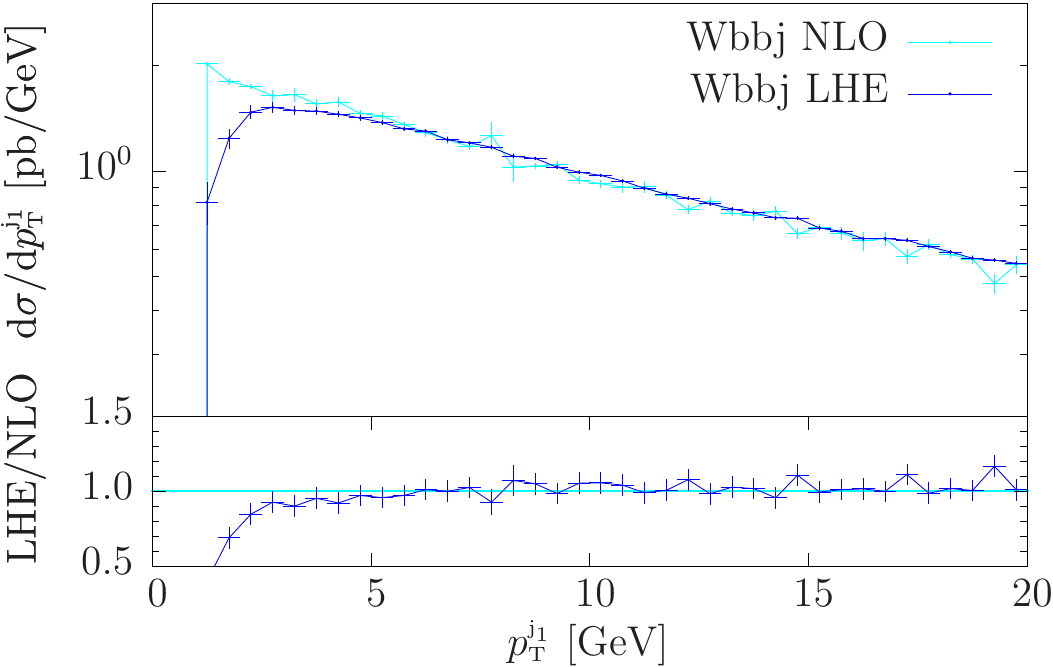}
\end{center}
\caption{Transverse-momentum distribution of the hardest jet at NLO and LHE
  level.}
\label{fig:WBBJ-NLO-LHE_j1-pt-001-000} 
\end{figure}
\begin{figure}[htb]
\begin{center}
\includegraphics[width=0.49\textwidth]{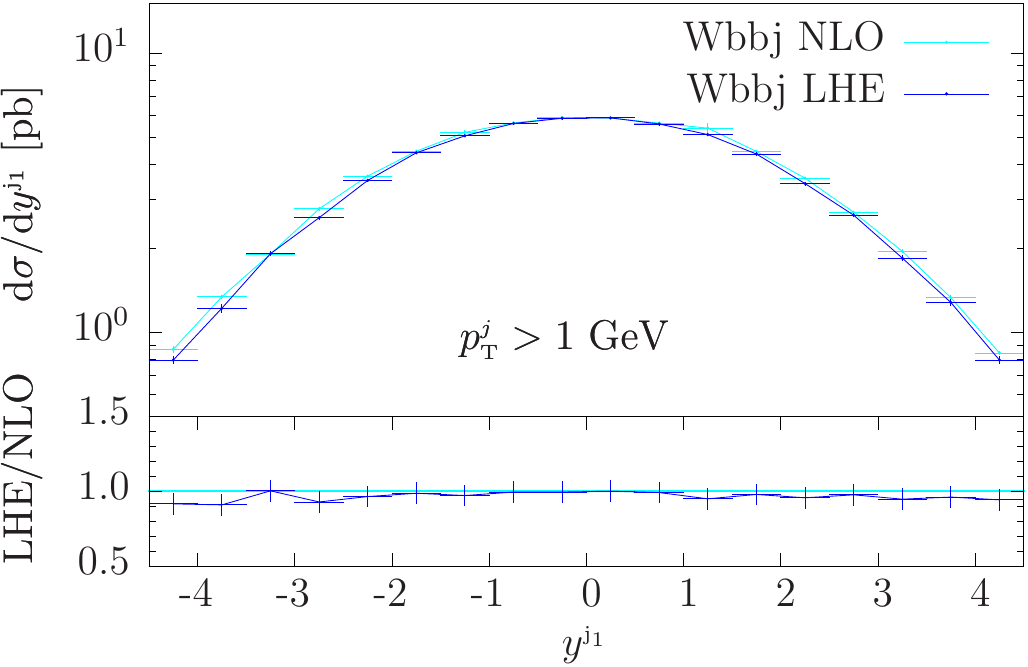}
\includegraphics[width=0.49\textwidth]{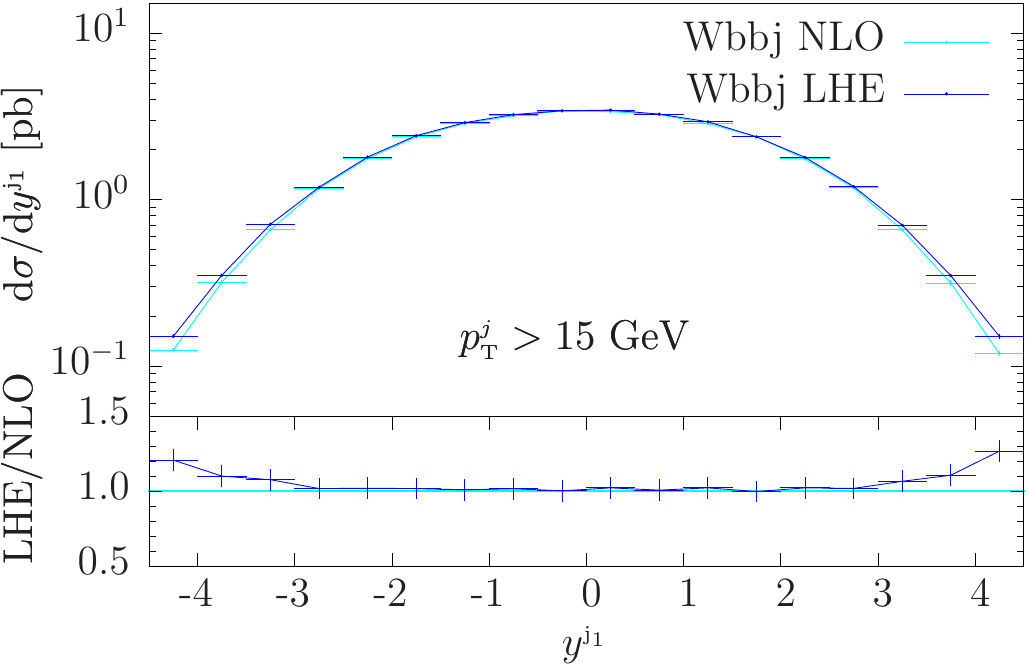}
\end{center}
\caption{Rapidity distribution of the hardest jet at NLO and LHE
  level.}
\label{fig:WBBJ-NLO-LHE_j1-y-015-000} 
\end{figure}
We first compare the transverse momentum and rapidity of the first hardest
jet, that are predicted by the \Wbbj{} generator with next-to-leading order
accuracy. 

In fig.~\ref{fig:WBBJ-NLO-LHE_j1-pt-001-000}, we plot the transverse momentum
of the hardest jet $\pt^{\sss\rm j_1}$ at NLO and LHE level. The agreement is
very good over a wide kinematic range.  In the right panel, the low
$\pt^{\sss\rm j_1}$ region is illustrated: here, the LHE distribution is
finite and goes to zero due to the Sudakov form factor coming from \MiNLO{},
applied to the primary process, and to the suppression factor associated to
the produced radiation, i.e.~the second jet, coming from the POWHEG Sudakov
form factor. In the NLO result, the latter Sudakov form factor is absent, and
the result increases.
 
We find very good agreement also for the rapidity of the hardest jet,
$y^{\sss \rm j_1}$, shown in fig.~\ref{fig:WBBJ-NLO-LHE_j1-y-015-000}.  In
the right panel, a minimum $\pt$ cut of 15~GeV on jets has been imposed.

\begin{figure}[htb]
\begin{center}
\includegraphics[width=0.49\textwidth]{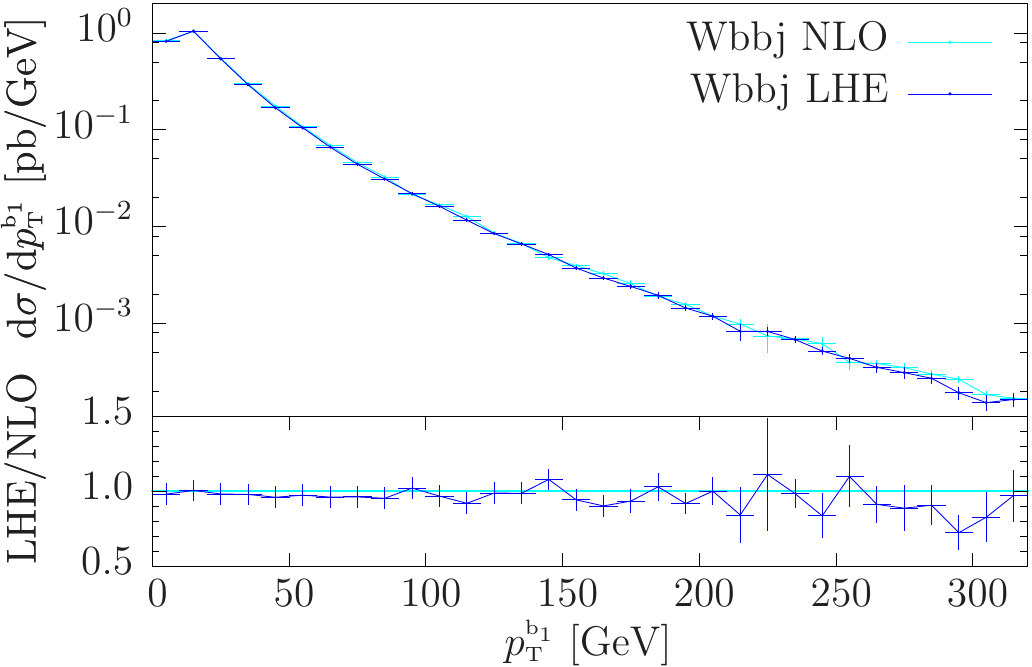}
\includegraphics[width=0.49\textwidth]{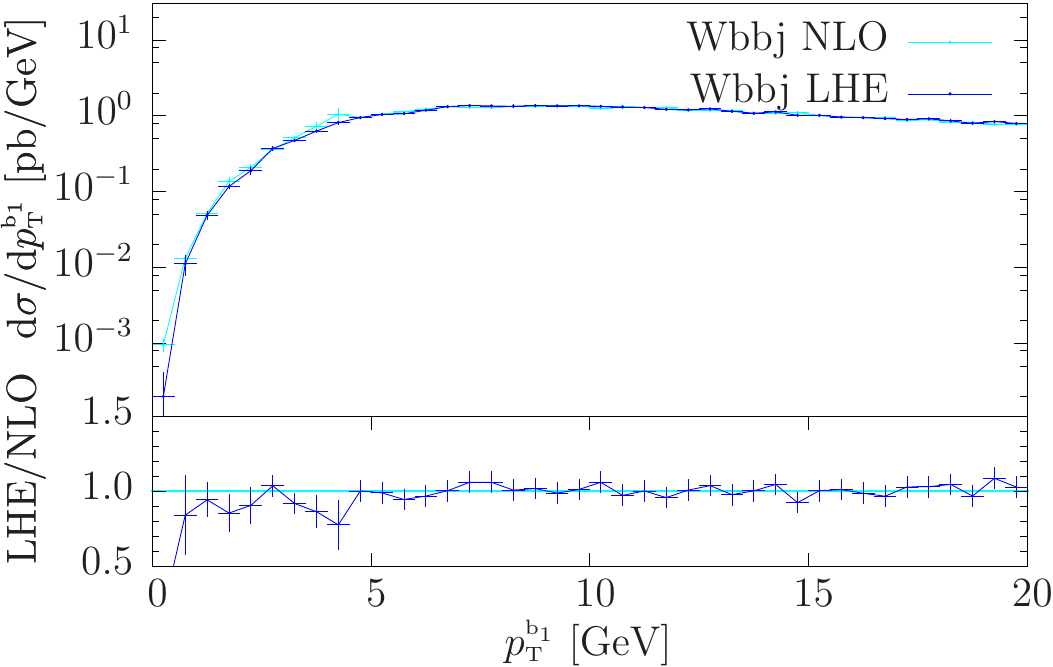}
\end{center}
\caption{Transverse-momentum distribution of the hardest $b$ jet.}
\label{fig:WBBJ-NLO-LHE_b1-pt-001-000} 
\end{figure}

\begin{figure}[htb]
\begin{center}
\includegraphics[width=0.49\textwidth]{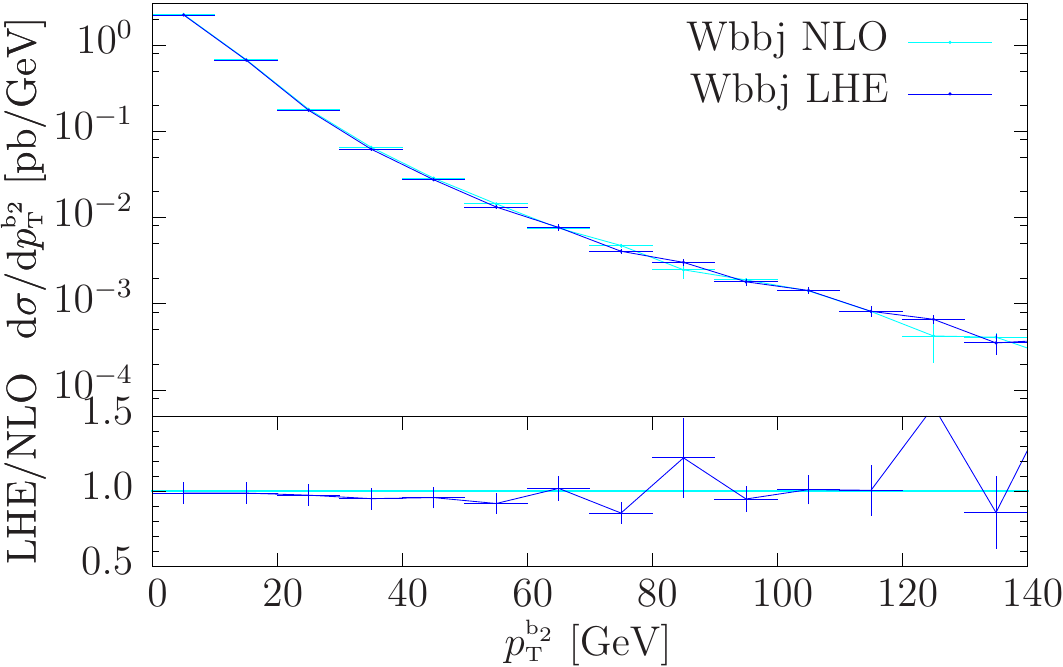}
\includegraphics[width=0.49\textwidth]{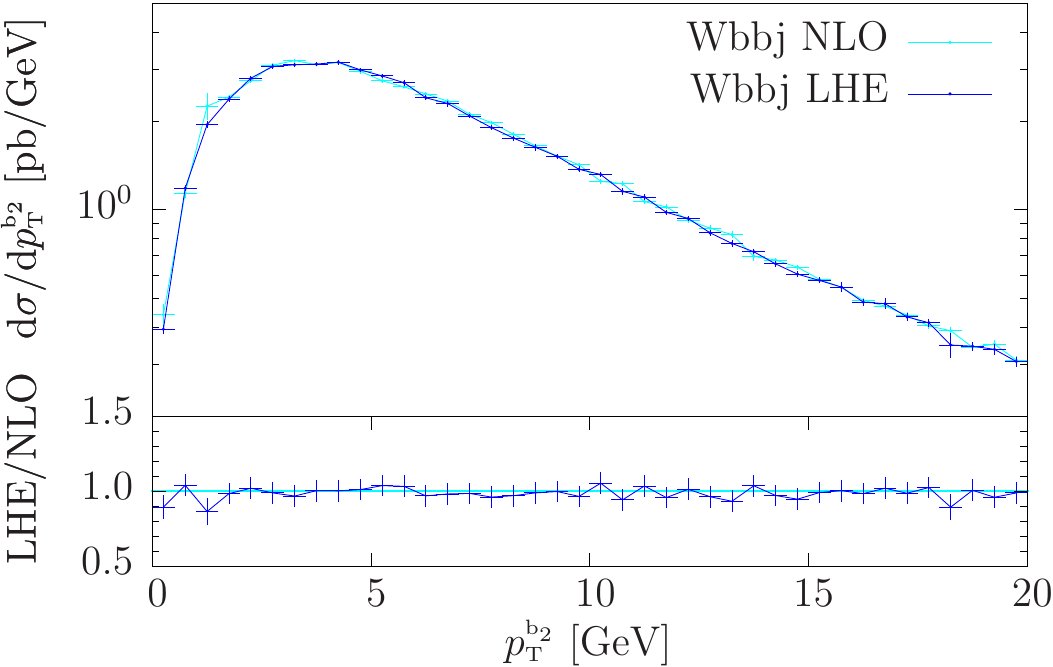}
\end{center}
\caption{Transverse-momentum distribution of the next-to-hardest $b$ jet.}
\label{fig:WBBJ-NLO-LHE_b2-pt-001-000}
\end{figure}
In figs.~\ref{fig:WBBJ-NLO-LHE_b1-pt-001-000}
and~\ref{fig:WBBJ-NLO-LHE_b2-pt-001-000}  we show the transverse-momentum
distribution for the hardest and next-to-hardest $b$ jets. In both cases, the
agreement between the NLO and LHE result is at the level of a few percent,
over the entire $\pt$ range. 
\begin{figure}[htb]
\begin{center}
\includegraphics[width=0.49\textwidth]{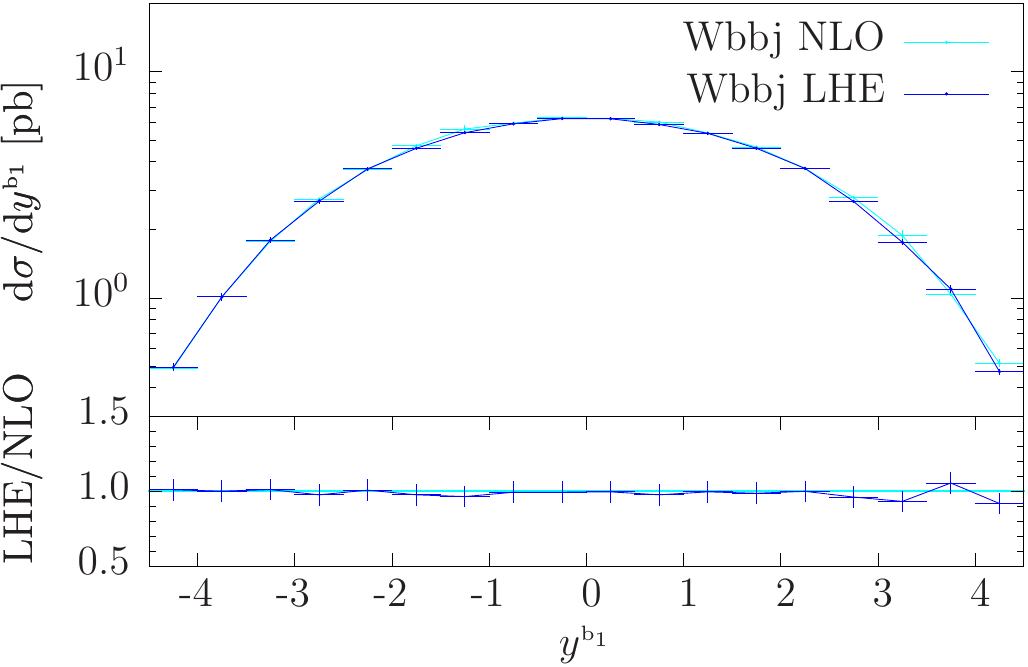}
\includegraphics[width=0.49\textwidth]{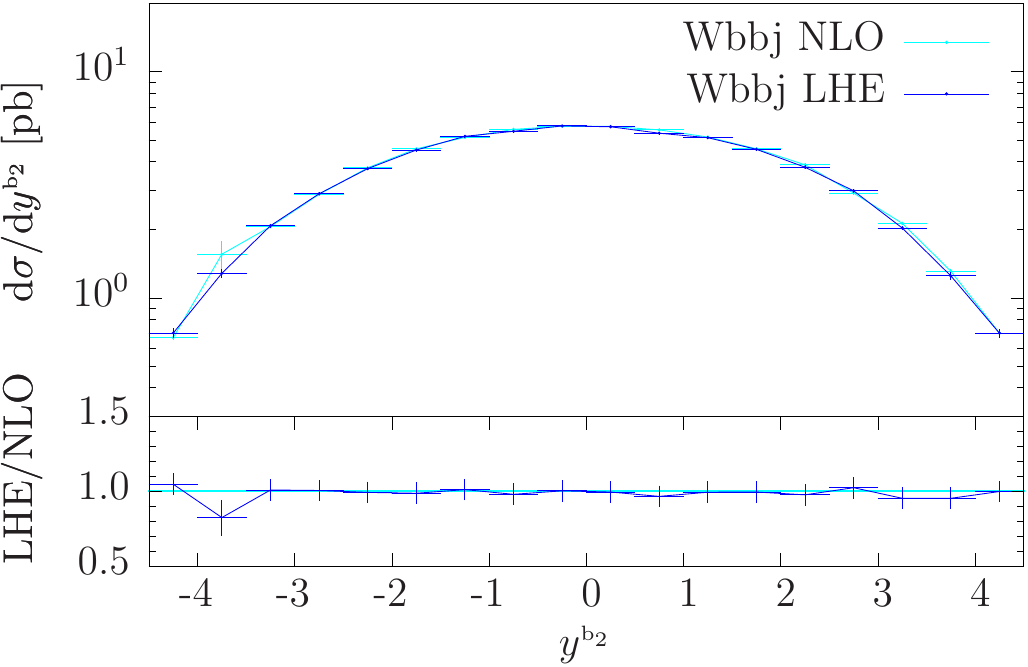}
\end{center}
\caption{Rapidity distribution of the hardest~(left panel) and
  next-to-hardest $b$ jet~(right panel). }
\label{fig:WBBJ-NLO-LHE_b1-y-001-000} 
\end{figure}
Similar conclusions hold for their rapidities, $y^{\sss \rm b_1}$ and $y^{\sss \rm
  b_2}$, illustrated in fig.~\ref{fig:WBBJ-NLO-LHE_b1-y-001-000}.

\begin{figure}[htb]
\begin{center}
\includegraphics[width=0.49\textwidth]{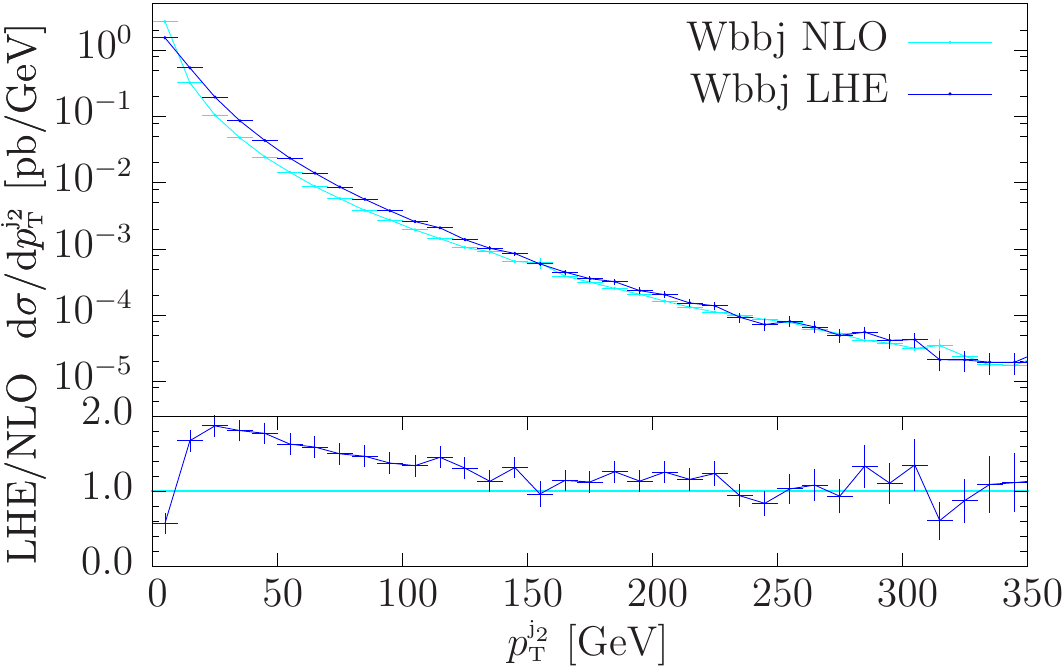}
\includegraphics[width=0.49\textwidth]{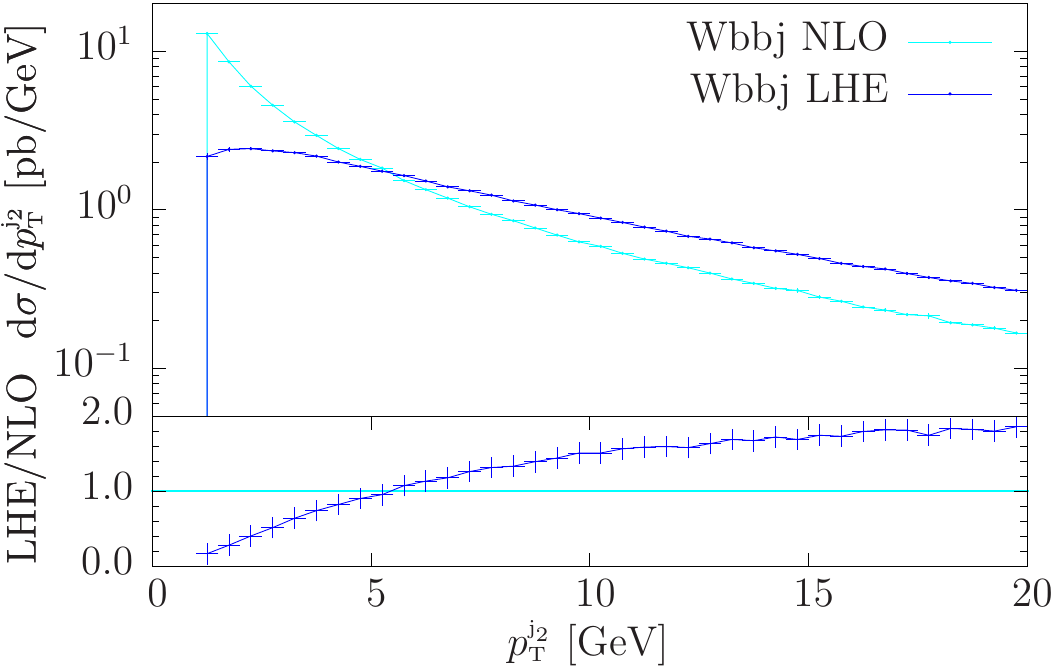}
\end{center}
\caption{Transverse-momentum distribution of the  next-to-hardest jet.}
\label{fig:WBBJ-NLO-LHE_j2-pt-001-000}
\end{figure}
The transverse-momentum distribution of the  next-to-hardest jet shown in
fig.~\ref{fig:WBBJ-NLO-LHE_j2-pt-001-000} is predicted at leading-order
only. The divergent behavior at small transverse momentum in the NLO result
is clearly visible in the right panel, where, instead, the POWHEG Sudakov form
factor damps the LHE result.

\begin{figure}[htb]
\begin{center}
\includegraphics[width=0.49\textwidth]{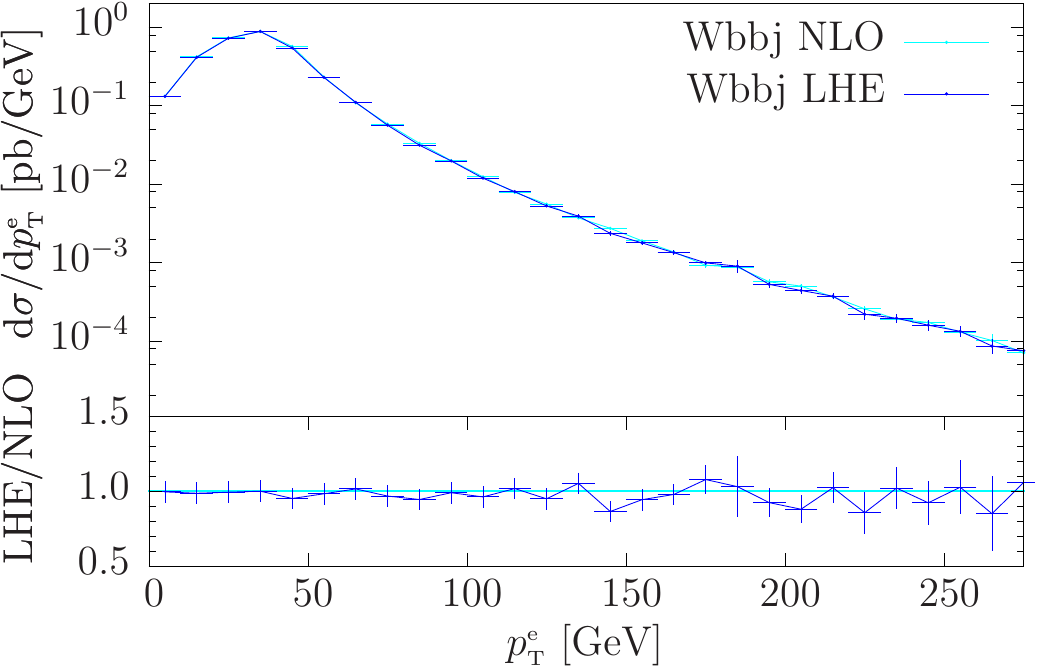}
\includegraphics[width=0.49\textwidth]{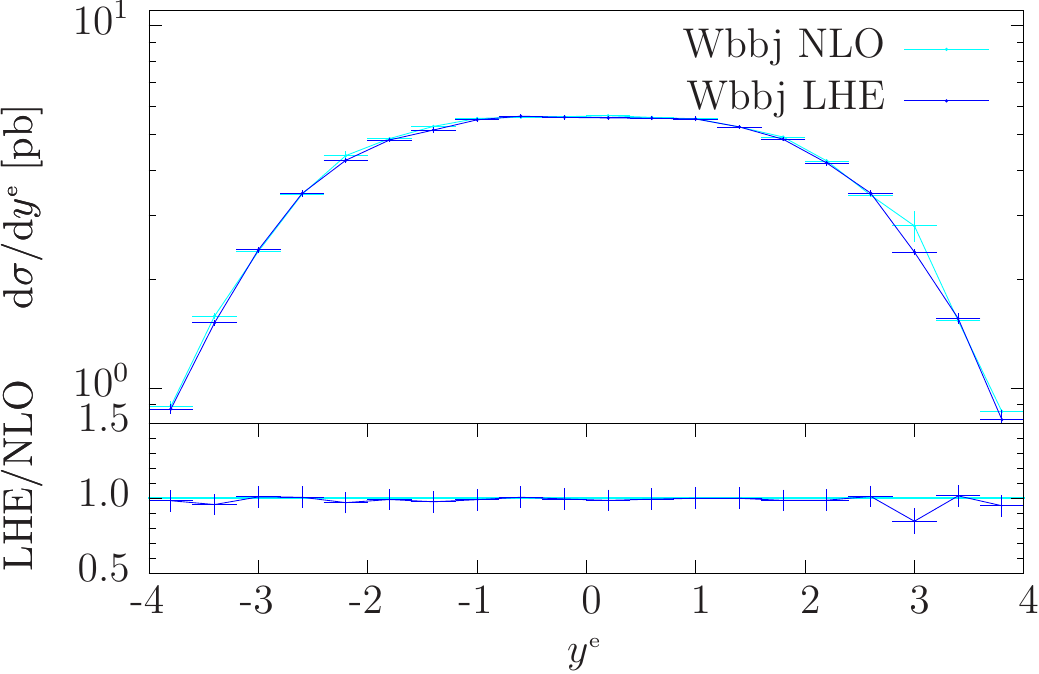}
\end{center}
\caption{Transverse-momentum~(left panel) and rapidity distribution~(right
  panel) of the charged lepton from $W$-boson decay.}
\label{fig:WBBJ-NLO-LHE_lept-pt-001-000}
\end{figure}
As last example, we plot the transverse-momentum and rapidity distribution of
the charged lepton from $W$-boson decay in
fig.~\ref{fig:WBBJ-NLO-LHE_lept-pt-001-000}, finding again very good
agreement between the NLO and LHE level results.

We have compared several other kinematic distributions with different cuts
and we have always found, when expected, excellent agreement between the NLO
and the LHE results.

\subsection{\Wbb{} and \Wbbj{}+\MiNLO{} comparisons}
\label{sec:Wbb_Wbbj_comparisons}
In this section we compare key distributions for $Wb\bar{b}$ and $Wb\bar{b}j$
production, computed using the \Wbb{} and \Wbbj{}+\MiNLO{} generators,
respectively, in order to investigate the level of agreement
between them, when the hardest jet $j$ becomes soft or collinear.

\begin{figure}[htb]
\begin{center}
\includegraphics[width=0.49\textwidth]{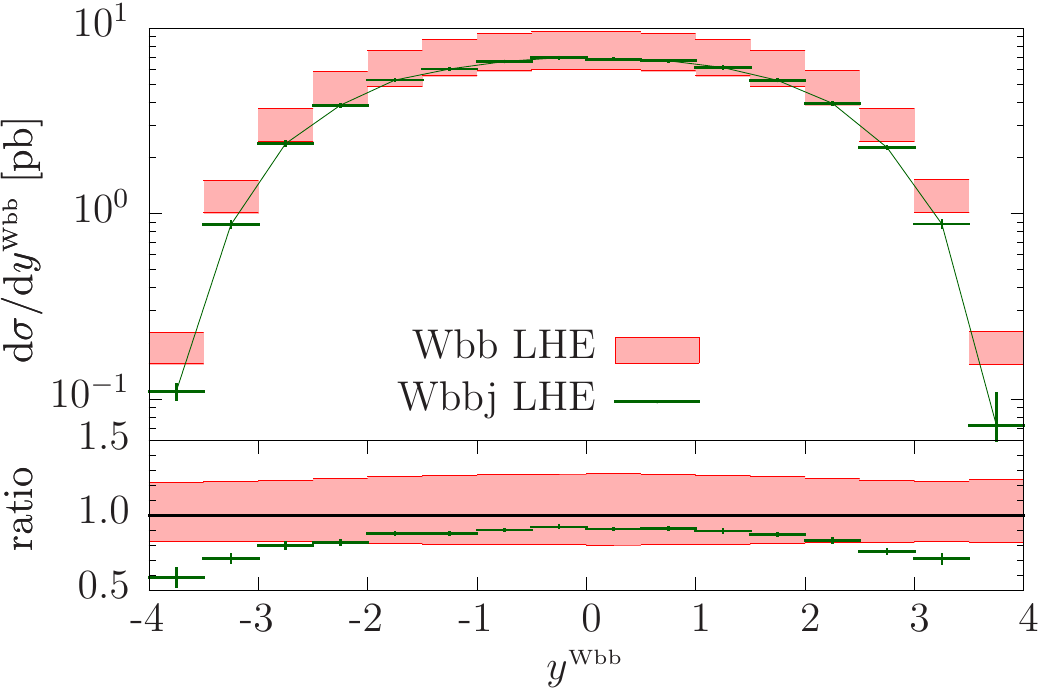}
\includegraphics[width=0.49\textwidth]{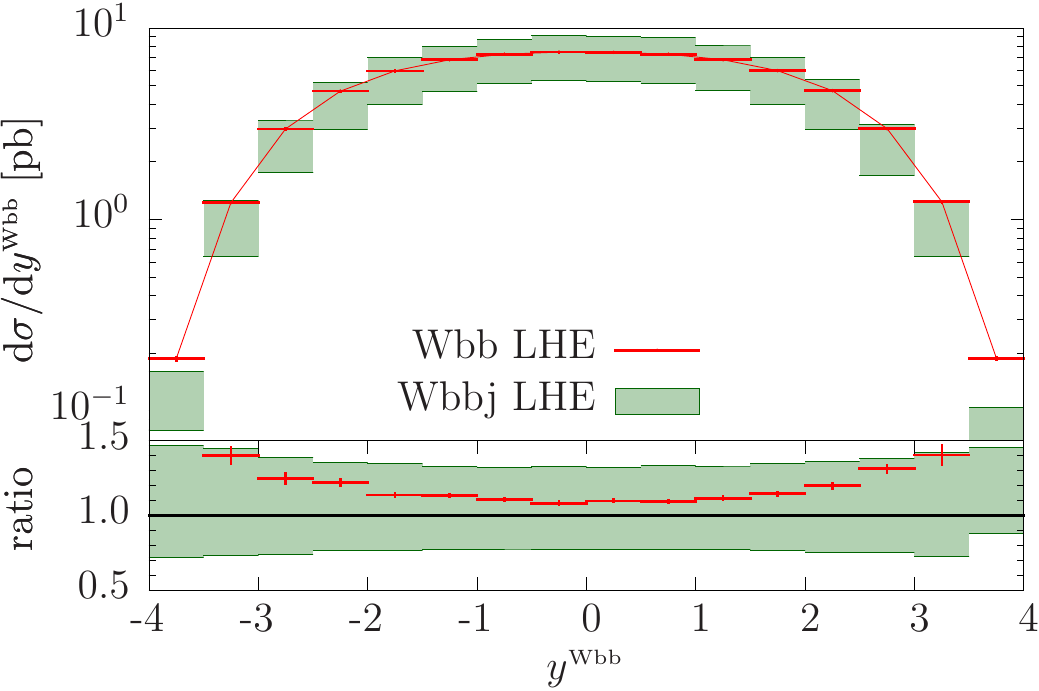}
\end{center}
\caption{Rapidity distribution of the $Wb\bar{b}$ system computed with the
  \Wbb{} and \Wbbj{}+\MiNLO{} generators, at the LHE level. In the left panel,
  we show the scale variation of the \Wbb{} generator in red, and the central
  value of the \Wbbj{} generator in green.  In the right panel, we plot the
  scale variation of the \Wbbj{}+\MiNLO{} code in green, and the central
  value of the \Wbb{} code in red.  In the lower insert, we display the ratio
  between the maximum and the minimum of the band over the respective central
  value.}
\label{fig:WBB-WBBJ_band-LHE_Wbb-y-001-000} 
\end{figure}
The rapidity $\yWbb$ of the $Wb\bar{b}$ system computed by the \Wbb{} generator
has NLO accuracy. In fact, this distribution receives contributions both at
the Born level and from the virtual and real diagrams.  In
fig.~\ref{fig:WBB-WBBJ_band-LHE_Wbb-y-001-000} we plot such quantity,
together with the same distribution as predicted by \Wbbj{}+\MiNLO{}.

The bands in the plots of this section are the envelope of the distributions
obtained by varying the renormalization and factorization scales by a factor
of 2 around the reference scale $\mu$ of eq.~(\ref{eq:Wbb_scales}), i.e.~by
multiplying the factorization  and the renormalization scale by the
scale factors $\KFA$ and $\KRA$, respectively, where
\begin{equation}
\label{eq:KRA_KFA}
(\KRA,\KFA)=(0.5,0.5),  (0.5,1), (1,0.5), (1,1),(2,1),(1,2),(2,2).
\end{equation}
These variations have been computed using the \POWHEGBOX{} reweighting
procedure, that recomputes the weight associated with an event in a fast way.

In the left panel of fig.~\ref{fig:WBB-WBBJ_band-LHE_Wbb-y-001-000}, we show
the scale variation of the \Wbb{} generator in red, and the central value of
the \Wbbj{} generator in green, at the LHE level.  In the right panel, we
plot the scale variation of the \Wbbj{}+\MiNLO{} code in green, and the
central value of the \Wbb{} code in red. In the lower insert, we display the
ratio between the maximum and the minimum of the band over the respective
central value.

The agreement between the predictions of the \Wbb{} and \Wbbj{}+\MiNLO{}
generators is within the scale-variation bands. We remind the reader that, if
we had used the \POWHEGBOX{} result for $Wb\bar{b}j$ production without
\MINLO, we could have not compare these distributions, since the rapidity of
the $Wb\bar{b}$ system would have been divergent in the limit of the
accompanying jet becoming soft or collinear with the incoming beams.

\begin{figure}[htb]
\begin{center}
\includegraphics[width=0.49\textwidth]{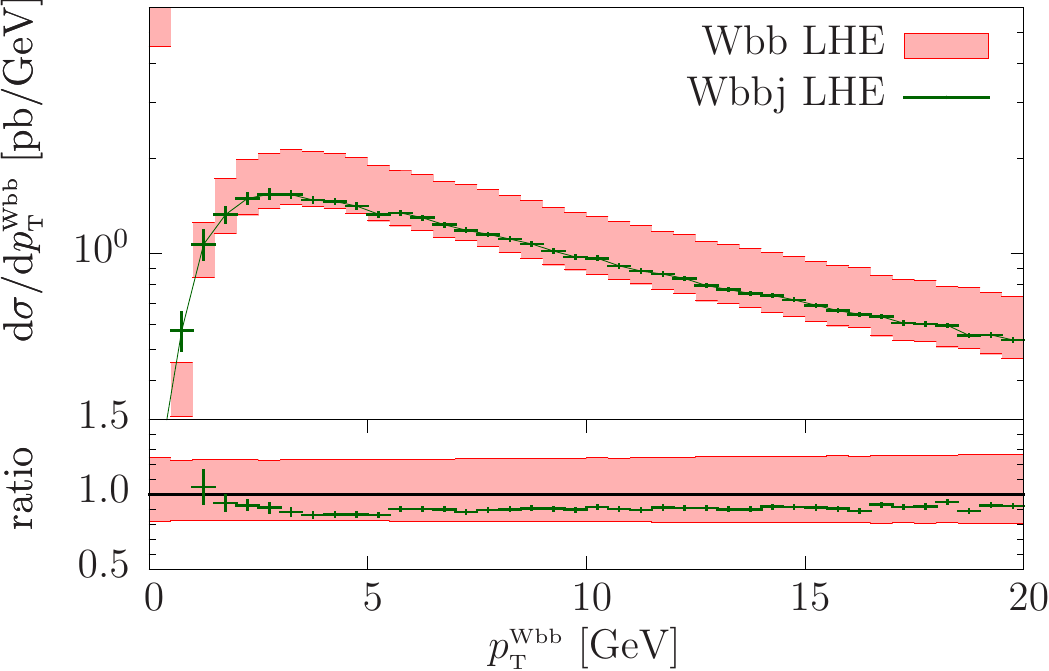}
\includegraphics[width=0.49\textwidth]{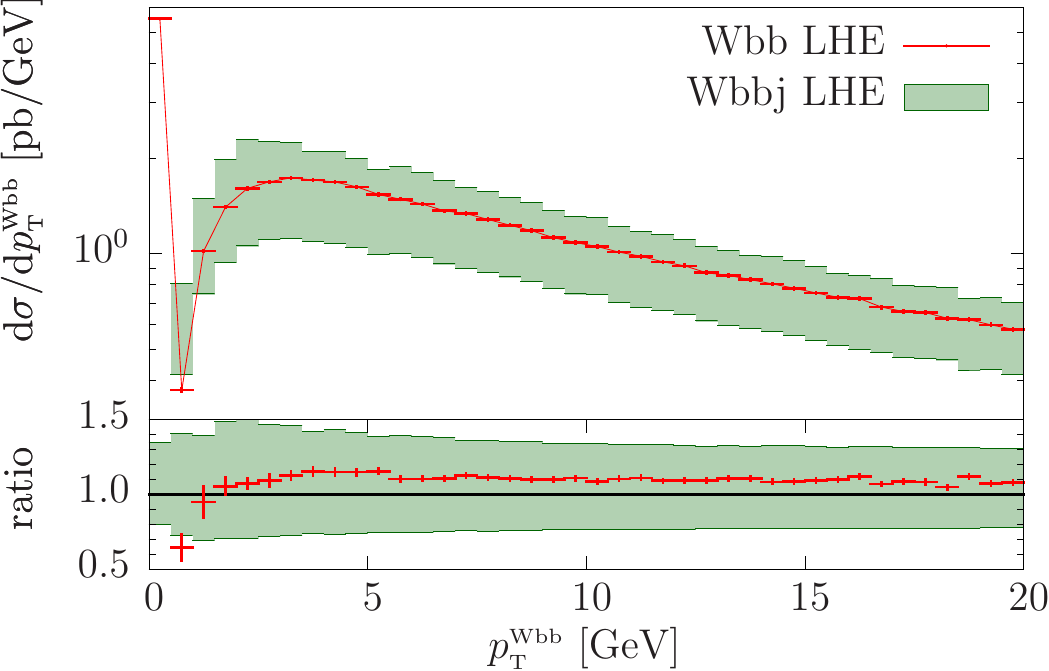}
\end{center}
\caption{Same as fig.~\ref{fig:WBB-WBBJ_band-LHE_Wbb-y-001-000} but for the
  transverse momentum of the $Wb\bar{b}$ system.}
\label{fig:WBB-WBBJ_band-LHE_Wbb-ptzoom2-001-000} 
\end{figure}
To better illustrate the behavior of the differential cross section in the
small transverse-momentum region, we compare the $\pt$ of the $Wb\bar{b}$
system obtained with the two generators in
fig.~\ref{fig:WBB-WBBJ_band-LHE_Wbb-ptzoom2-001-000}.  In $Wb\bar{b}$
production, this distribution is predicted with leading-order accuracy and
the POWHEG Sudakov form factor attached to the radiation makes it finite in
the small-$\pt$ region (the $Wb\bar{b}$ system recoils against the only hard
jet generated by the \POWHEGBOX).  In $Wb\bar{b}j$ production, this
distribution is finite due to the presence of the POWHEG Sudakov form factor
attached to the radiation, most likely the next-to-hardest jet, and to the
\MINLO{} Sudakov form factor, attached to the hardest jet accompanying the
$Wb\bar{b}$ system.  Again the agreement is very good. The finite
contribution to the differential cross section visible in the first $\pTWbb$
bin, in $Wb\bar{b}$ production, is due to events that have not radiated at the
LHE level.
\begin{figure}[htb]
\begin{center}
\includegraphics[width=0.49\textwidth]{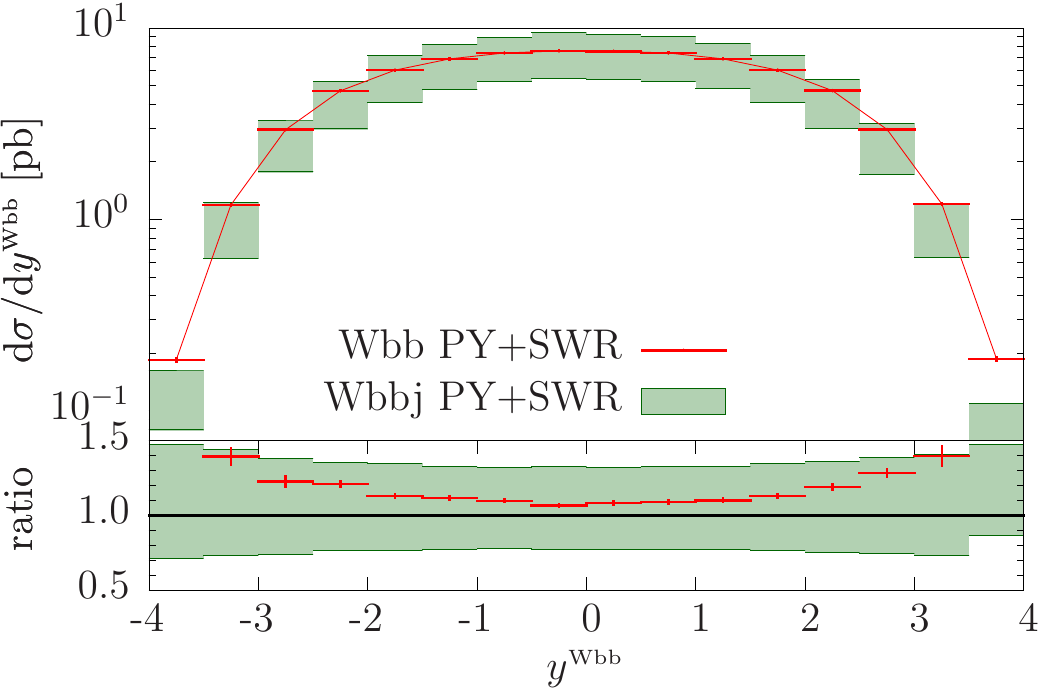}
\includegraphics[width=0.49\textwidth]{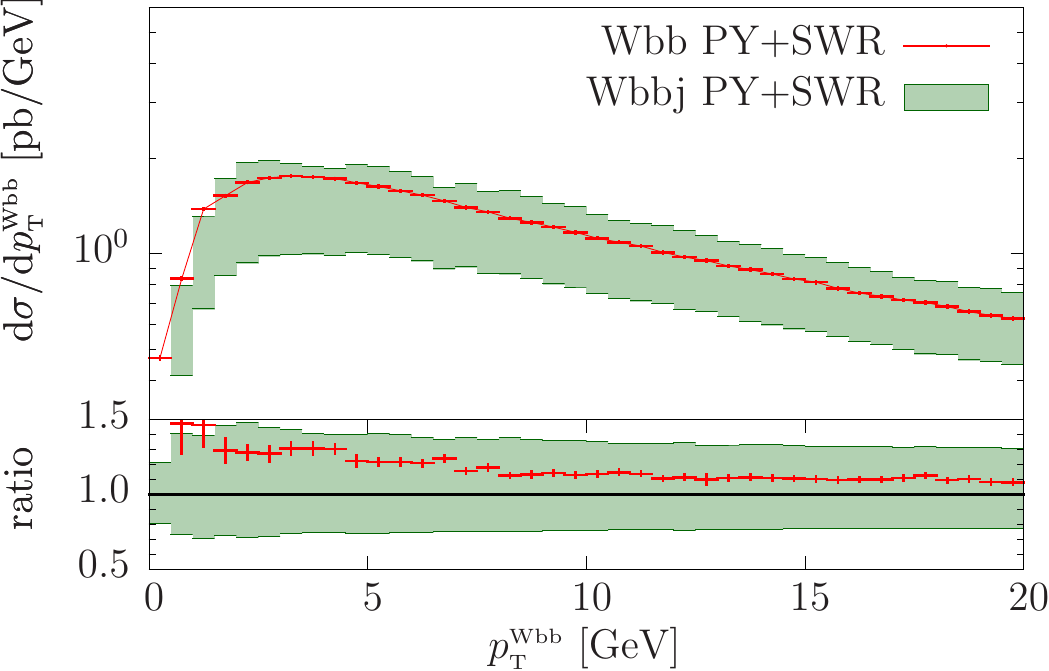}
\end{center}
\caption{Rapidity and transverse momentum distribution of the $Wb\bar{b}$
  system, generated by the \Wbbj+\MINLO{} code, and showered by \PYTHIA.}
\label{fig:WBB-WBBJ_band-SWR_Wbb-y-001-000}
\end{figure}
This peak is diluted away when the whole shower is completed by a Monte Carlo
program such as \PYTHIA{} or \HERWIG~\cite{Corcella:2000bw,Corcella:2002jc}, as
shown in fig.~\ref{fig:WBB-WBBJ_band-SWR_Wbb-y-001-000}, right panel.  In
this figure, we plot the differential cross sections as a function of $\yWbb$
and $\pTWbb$ after the shower has been completed by \PYTHIA. No
hadronization has been switched on at this level, but we have explicitly
checked that it has a negligible effect on these distributions.

\begin{figure}[htb]
\begin{center}
\includegraphics[width=0.49\textwidth]{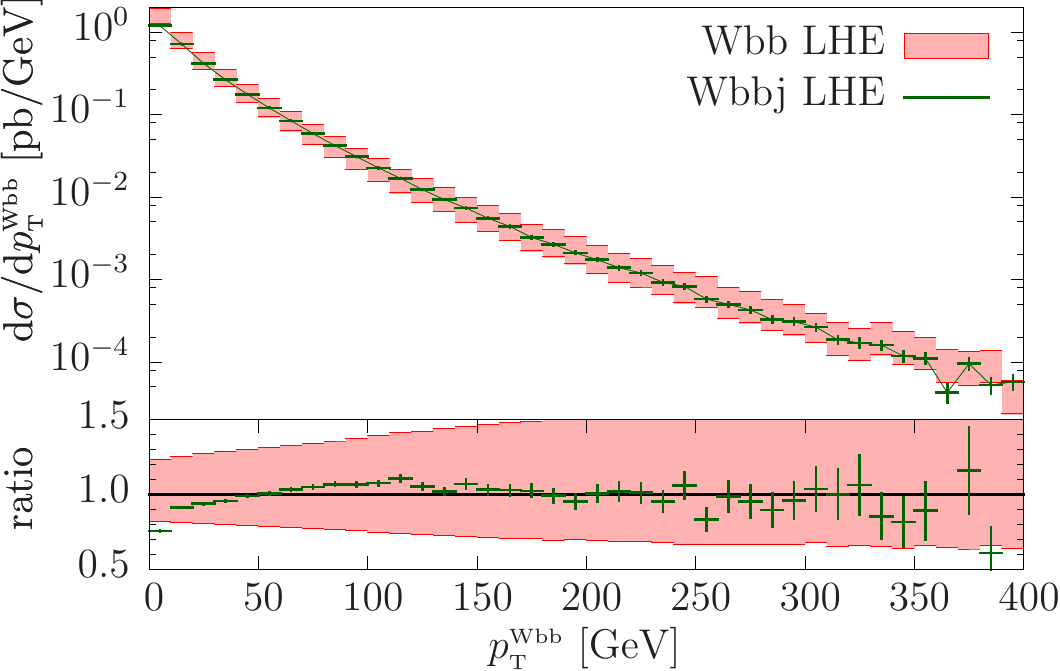}
\includegraphics[width=0.49\textwidth]{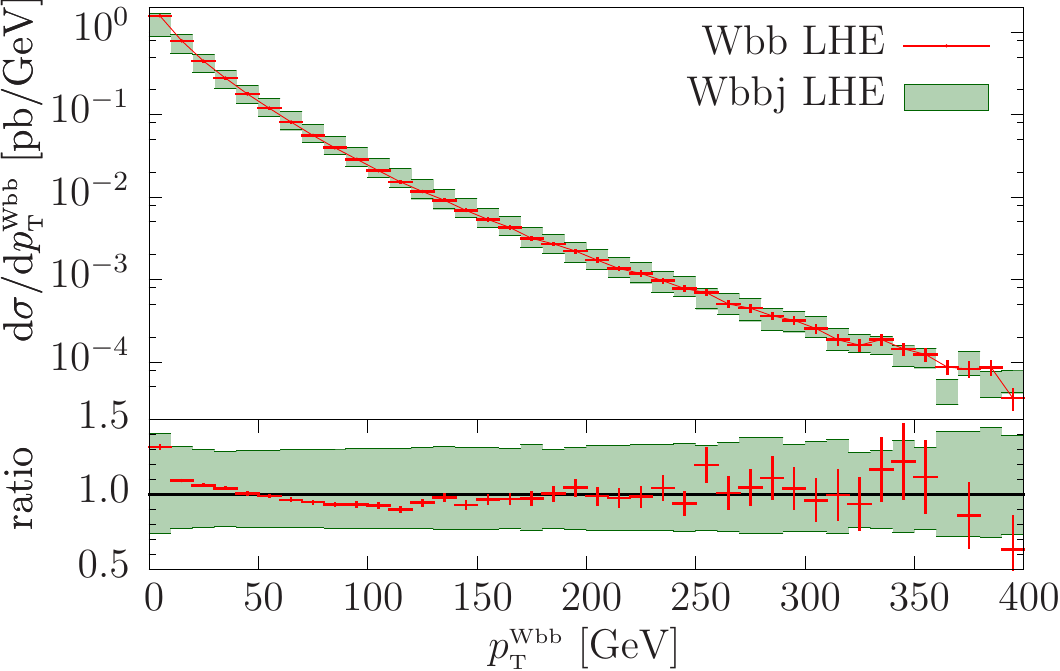}
\end{center}
\caption{Same as fig.~\ref{fig:WBB-WBBJ_band-LHE_Wbb-y-001-000} but for the
  transverse momentum of the $Wb\bar{b}$ system.}
\label{fig:WBB-WBBJ_band-LHE_Wbb-pt-001-000}
\end{figure}
Away from the small transverse-momentum region, the differential cross
section as a function of $\pTWbb$ is predicted at LO by the \Wbb{} generator,
and at NLO by the \Wbbj{} one. This is clearly displayed in
fig.~\ref{fig:WBB-WBBJ_band-LHE_Wbb-pt-001-000}, where the scale variation
bands for $Wb\bar{b}$ production reaches the 70\%, while they are around 40\%
for $Wb\bar{b}j$ production.

\section{Comparison with ATLAS and CMS data}
\label{sec:data}
In this section we compare the results that we obtained with the \Wbb{} and
\Wbbj{} generators with the available experimental data.

\begin{figure}[htb]
\begin{center}
\includegraphics[width=0.49\textwidth]{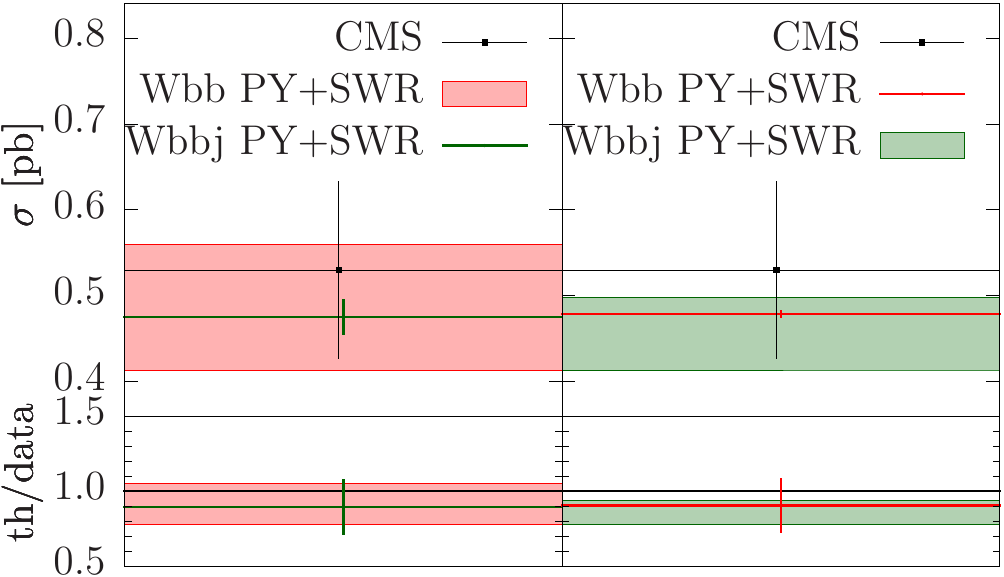}
\includegraphics[width=0.49\textwidth]{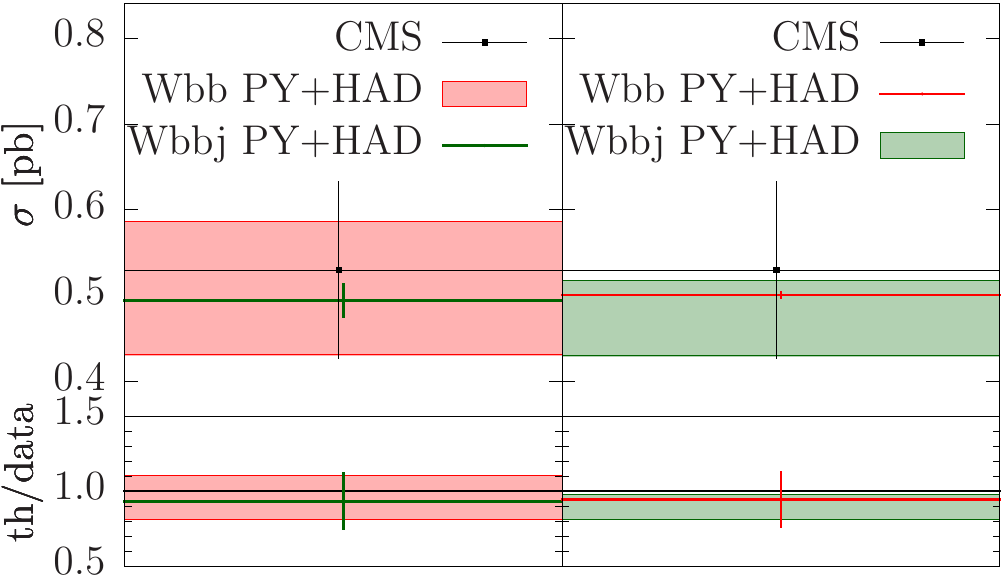}
\end{center}
\caption{Total cross section for $Wb\bar{b}$ production within the cuts of
  eq.~(\ref{eq:CMS_cuts}). In the left panel, the results after the shower
  done by \PYTHIA{}. In the right plot, the full showered and hadronized results
  are shown. The measured value is compared with the results from the \Wbb{}
  and \Wbbj+\MINLO{} generators.}
\label{fig:WBB-WBBJ_band-CMS-SWR_XS-Wbb}
\end{figure}
The CMS Collaboration has measured the $Wb\bar{b}$ cross section at the LHC
at 7~TeV and the result, reported in ref.~\cite{Chatrchyan:2013uza}, is 
\begin{equation}
\sigma(pp \to W b \bar{b} )\times {\rm BR}(W\to \mu \nu)= 0.53 \pm 0.10~{\rm
  pb}\,, 
\end{equation} 
within the following experimental cuts on jets and charged
leptons 
\begin{equation}
\label{eq:CMS_cuts}
\pt^{\sss\rm j} > 25~{\rm GeV}\,, \quad \left|y^{\rm\sss
  j}\right|<2.4\,, \quad \pt^{\rm \sss e} > 25~{\rm GeV}\,,\quad \left|y^{\rm
 \sss e}\right|<2.1\,, \quad \Delta R_{\rm\sss j,e} > 0.5 \,.  
\end{equation} 
To reconstruct jets the anti-$\kt$ algorithm with $R=0.5$ was used and
only events with exactly two jets which passed the $b$-tagging requirements
were taken into account.  In fig.~\ref{fig:WBB-WBBJ_band-CMS-SWR_XS-Wbb} we
compare our predictions with the measured value. In the left panel, we show
our result after the shower done by \PYTHIA, and in the right panel, the same
result at the hadronic level. In particular, with the \Wbb{} generator we have
\begin{equation}
\sigma(pp \to W b \bar{b} )\times {\rm BR}(W\to \ell \nu_\ell)=
0.50\,^{+0.09}_{-0.07}~{\rm pb} 
\end{equation}
and with the \Wbbj+\MINLO{} one
\begin{equation} 
\sigma(pp \to W b \bar{b} )\times {\rm BR}(W\to \ell \nu_\ell) = 0.49\,^{+0.03}_{-0.06}~{\rm pb}
\end{equation}
at the hadronic level.  We find very good agreement between the cross section
computed with the \Wbb{} generator and the \Wbbj+\MINLO{} one, and both are
consistent with the measured data.

\begin{figure}[htb]
\begin{center}
\includegraphics[width=0.49\textwidth]{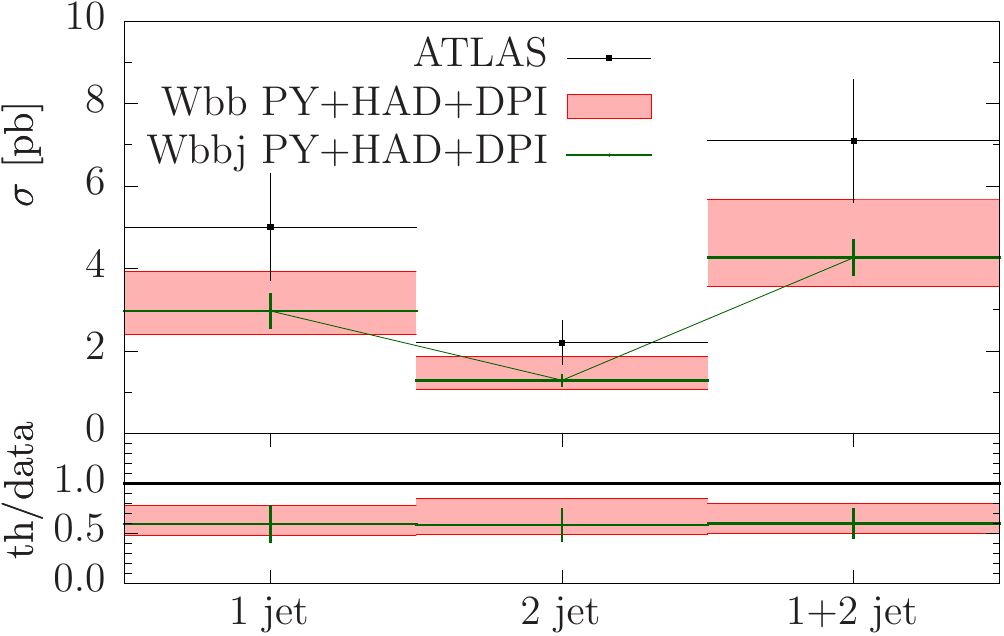}
\includegraphics[width=0.49\textwidth]{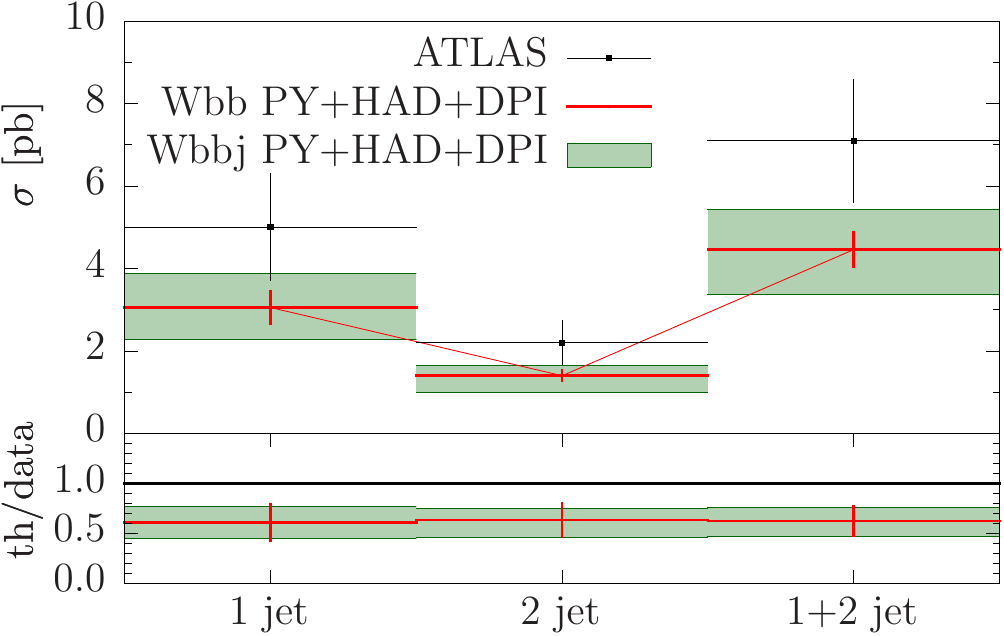}
\end{center}
\caption{Cross sections, within the cuts described in the text, for the ATLAS
  measurement of ``1 jet'', ``2 jet'' and ``1+2 jet'' $Wb$ production. The
  theoretical central results and error bands have been corrected for the DPI
  effects of tab.~\ref{tab:DPI}. Results are at the full shower+hadron
  level.}
\label{fig:WBBJ_band-ATLAS-SWR_XS-Wbj-ATLAS2}
\end{figure}
The ATLAS Collaboration reported a measurement of $W$+ $b$-jets ($W+b+X$ and
$W+b\bar{b}+X$) cross section at 7~TeV in ref.~\cite{Aad:2013vka}.  Candidate
$W +b$-jets events are required to have exactly one high-$\pT$ electron or
muon, as well as missing transverse momentum consistent with a neutrino from
a $W$ boson, and one or two reconstructed jets, exactly one of which must be
$b$-tagged. Events with two or more $b$-tagged jets are rejected, as are
events with three or more jets.  The details of the analysis are as follows:
jets are reconstructed using the anti-$\kt$ algorithm, with a radius
parameter $R=0.4$, and are required to have a transverse momentum greater
than 25~GeV and absolute rapidity $|y^{\sss j}|< 2.1$. Furthermore the following
cuts are applied
\begin{equation}
\pt^{\sss \rm e} > 25~{\rm GeV}\,,\quad \pt^{\sss \rm \nu} > 25~{\rm GeV}\,,\quad
\left|y^{\sss \rm e}\right|<2.5\,, \quad m_{\sss \rm T}^{\sss \rm W^{\pm}} > 60~{\rm
  GeV}\,, \quad \Delta R_{\rm \sss j,e} > 0.5 \,.
\end{equation}
In fig.~\ref{fig:WBBJ_band-ATLAS-SWR_XS-Wbj-ATLAS2} we plot the ATLAS results
for the measured cross-sections for the ``1 jet'', ``2 jet'' and ``1+2 jet''
fiducial regions, together with our results. In the ``1 jet'' bin, we show
the cross section with only one jet, that necessarily must contain at least
one $b$ quark, or a $\bar{b}$, or both (clustered in a single jet).  In the
``2 jet'' bin, we plot the cross section for events with two jets, only one
of which is $b$-tagged. In the ``1+2 jet'' bin, there is the sum of the
previous two cross sections.

With the same scale choice of sec.~\ref{sec:scale_choice}, our predictions
for the ``1+2 jet'' bin are
\beqn
\sigma_\Wbbj({\rm \mbox{1+2~jet}}) &=& 2.93\, ^{+0.73}_{-0.59}~{\rm pb} \nonumber\\
\sigma_\Wbb({\rm \mbox{1+2~jet}}) &=&  3.12\,^{+0.81}_{-0.58}~{\rm pb}
\eeqn
at the hadron level. 
\begin{table}[htb]
\centering
\begin{tabular}{|c|c|c|c|}
\hline
correction & 1 jet & 2 jet & 1+2 jet \\
\hline
\phantom{\Big{|}}DPI~[pb] & $ \ \ 1.02\,^{+0.40}_{-0.29}$ \ \ &
$\ \  0.32\,^{+0.12}_{-0.09}\ \ $ & \ \ $1.34\,^{+0.42}_{-0.30}\ \ $ \\
\hline
 \end{tabular}
\caption{Additive corrections for double-parton interactions~(DPI) as estimated
  by the ATLAS Collaboration. See tab.~7 in ref.~\cite{Aad:2013vka}.}
\label{tab:DPI} 
\end{table}
Since neither the \Wbb{} nor the \Wbbj+\MINLO{} results contain the effect of
double-parton interactions~(DPI) within the same proton, we need to correct
our cross sections for this.  The computation of this contribution is beyond
the aim of the present paper. On the other hand, the ATLAS Collaboration has
estimated such effect and has provided some additive values to correct for
it.  We collect them in tab.~\ref{tab:DPI} and after correcting for~DPI, we
get
\begin{eqnarray}
\label{eq:sig_ATLAS_DPI}
\sigma_{\Wbbj+{\rm\sss DPI}}({\rm \mbox{1+2~jet}}) &=&  4.26\,^{+1.16}_{-0.89}~{\rm
  pb}
\\
\label{eq:sig_ATLAS_DPI_wbb}
\sigma_{\Wbb+{\rm\sss DPI}}({\rm \mbox{1+2~jet}}) &=&  4.46\,^{+0.97}_{-0.89}~{\rm pb}
\end{eqnarray}
where we have estimated the total errors by simply adding in quadrature the
errors from different sources.
These results, plus the ones for ``1 jet'' and ``2 jet'' are plotted in
fig.~\ref{fig:WBBJ_band-ATLAS-SWR_XS-Wbj-ATLAS2}. Our predictions in
eqs.~(\ref{eq:sig_ATLAS_DPI}) and~(\ref{eq:sig_ATLAS_DPI_wbb}) should be
compared with the measured value
\begin{equation}
\sigma({\rm \mbox{1+2~jet}}) = 7.1 \pm 1.5~{\rm pb}\,.
\end{equation}
Although the theoretical results and the experimental data are consistent
between each other, the level of agreement is not so good: in fact the error
bands overlap only marginally.

\begin{figure}[htb]
\begin{center}
\includegraphics[width=0.49\textwidth]{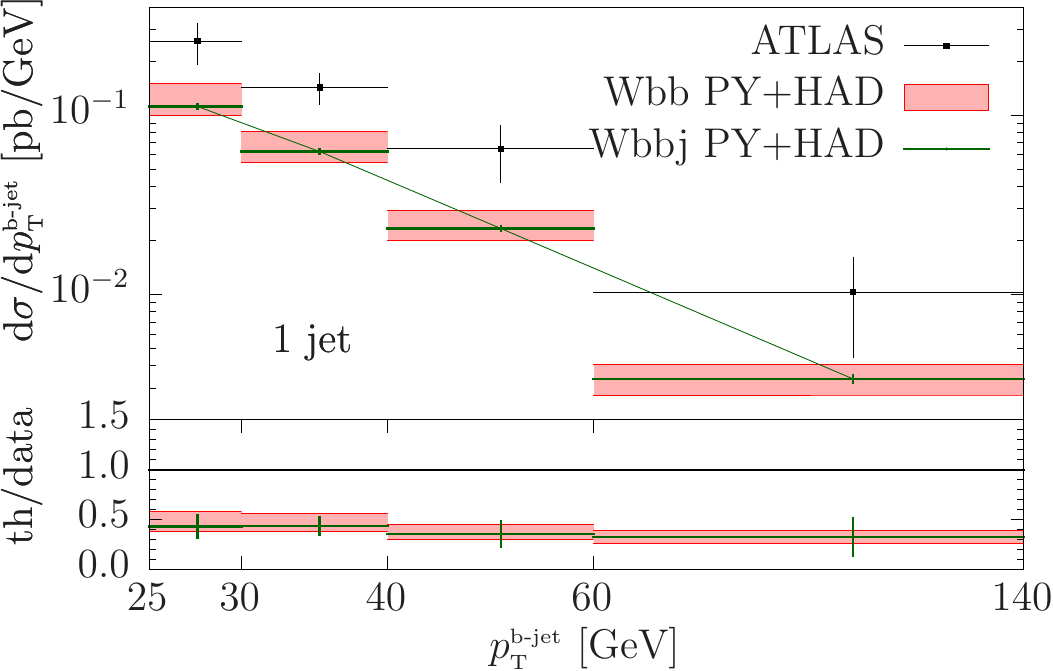}
\includegraphics[width=0.49\textwidth]{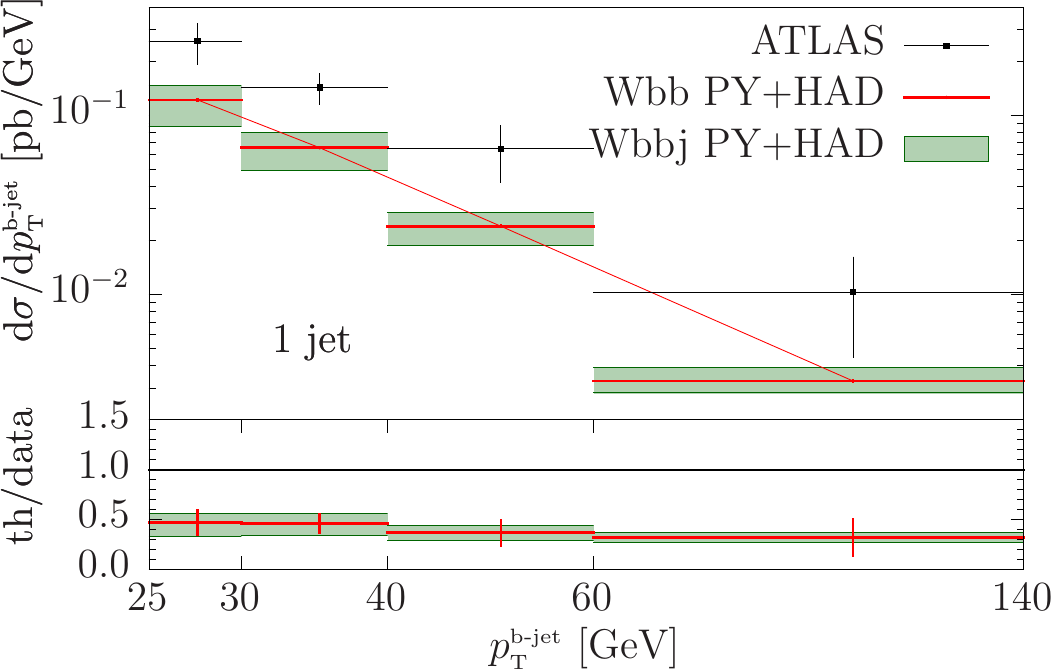}
\end{center}
\caption{Transverse-momentum distribution of the $b$-tagged jet in the  ``1
  jet'' sample. The theoretical results are at the shower+hadron level. No
  DPI corrections included.}
\label{fig:WBBJ_band-ATLAS-HAD_b-pt-1j}
\end{figure}

\begin{figure}[htb]
\begin{center}
\includegraphics[width=0.49\textwidth]{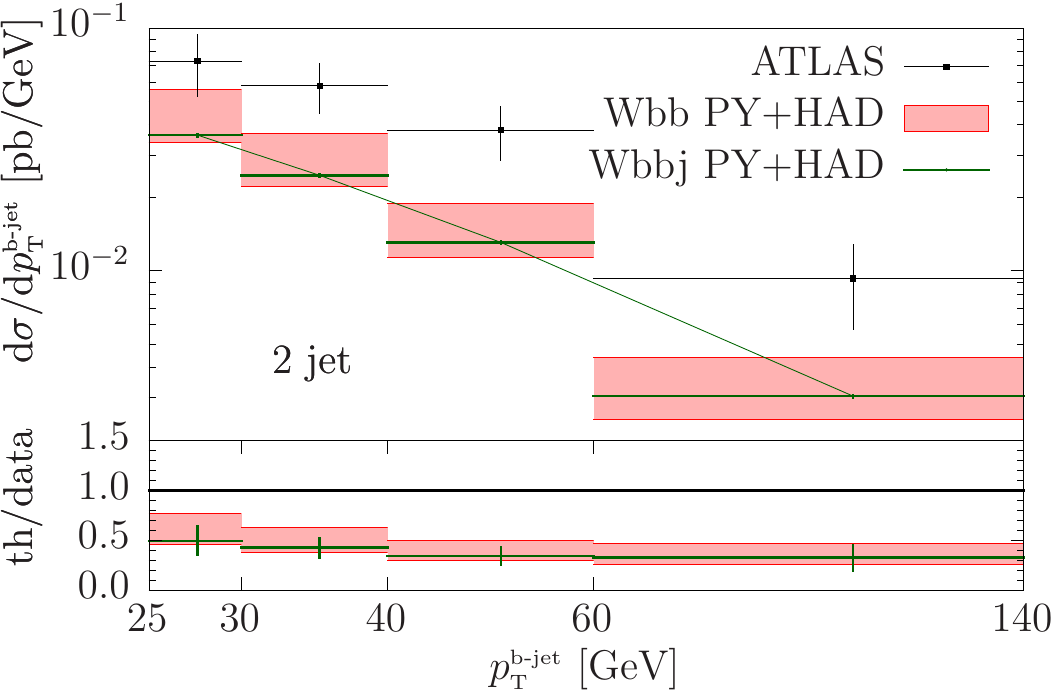}
\includegraphics[width=0.49\textwidth]{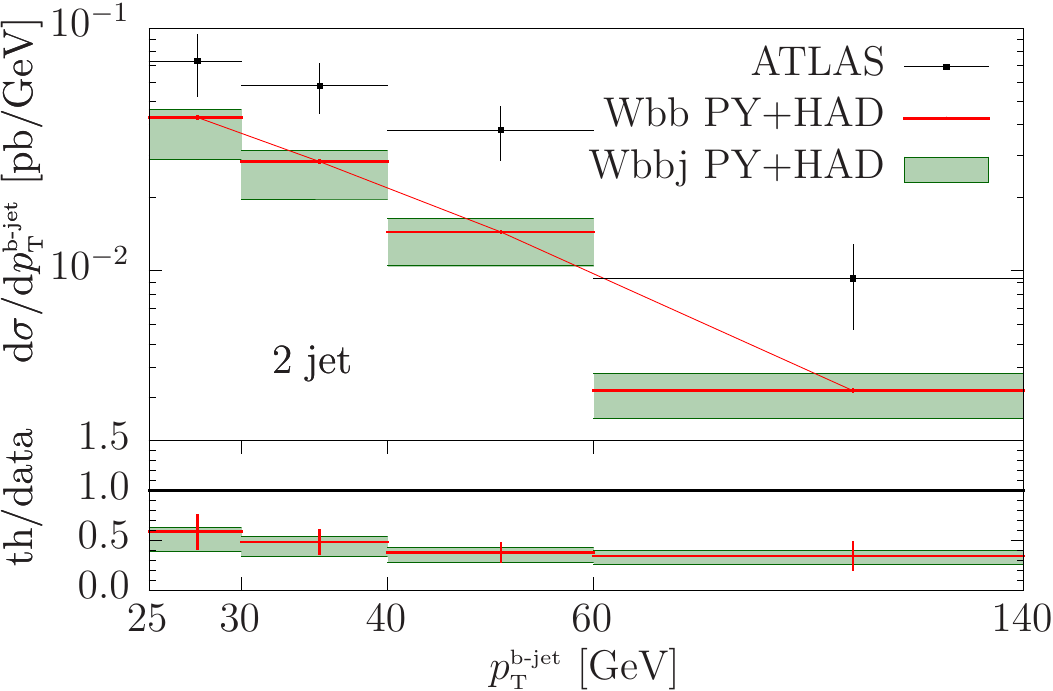}
\end{center}
\caption{Transverse-momentum distribution of the $b$-tagged jet in the  ``2
  jet'' sample. The theoretical results are at the shower+hadron level. No
  DPI corrections included.}
\label{fig:WBBJ_band-ATLAS-HAD_b-pt-2j}
\end{figure}

The ATLAS Collaboration has also measured the $\pt$ spectrum of the
$b$-tagged jet in the ``1 jet'' and ``2 jet'' samples. We plot the measured
values and our theoretical results in
figs.~\ref{fig:WBBJ_band-ATLAS-HAD_b-pt-1j}
and~\ref{fig:WBBJ_band-ATLAS-HAD_b-pt-2j}.  No DPI corrections are available
for these quantities, and this partially explains the discrepancy between
theory and data, both with the \Wbb{} and \Wbbj+\MINLO{} generator.

\subsection{A different scale choice}
All the results presented so far have been computed at the scales illustrated
in sec.~\ref{sec:scale_choice}. In this section, we show a few results at the
same scale used by the ATLAS Collaboration in ref.~\cite{Aad:2013vka}, where
the measured cross sections are compared with the predictions of mixed 4- and
5-flavour NLO calculations~\cite{Campbell:2006cu,Campbell:2008hh,
  Caola:2011pz}, computed using \MCFM~\cite{MCFM}. The results are generated
using the following central renormalization and factorization dynamical scale
\begin{equation}
\label{eq:ATLAS_scale}
\muf^2=\mur^2=\mu^2=m^2_{\sss\rm e\nu} +\(\pt^{\sss\rm e\nu}\)^2 +
\frac{\mb^2+\(\pt^{\sss\rm b}\)^2}{2}+
\frac{\mb^2+\(\pt^{\sss\rm \bar{b}}\)^2}{2}\,.
\end{equation}
The scale-variation bands computed in ref.~\cite{Aad:2013vka} are obtained by
varying $\mu$ between a quarter and four times the central value, while in
our results we have varied the scales independently as discussed in
sec.~\ref{sec:Wbb_Wbbj_comparisons}.

\begin{table}[htb]
\centering
\begin{tabular}{|c|c|c|c||c|c|}
\hline
 \phantom{\Big{|}}    & \MCFM{} NLO  & LHE & PY+SWR  & \MCFM+HAD & PY+HAD    \\
\hline
\phantom{\Big{|}}``1 jet''& $2.16\,^{+0.78}_{-0.59}$ & $2.54\,^{+0.27}_{-0.41}$ & $2.39\,^{+0.23}_{-0.37}$ & $1.99\,^{+0.72}_{-0.54}$ & $2.40\,^{+0.22}_{-0.37}$\\
\hline
\phantom{\Big{|}}``2 jet''& $1.43\,^{+0.42}_{-0.24}$ & $1.52\,^{+0.38}_{-0.30}$ & $1.24\,^{+0.31}_{-0.25}$ & $1.37\,^{+0.40}_{-0.23}$ & $1.27\,^{+0.31}_{-0.25}$         \\  
\hline
 \end{tabular}
\caption{Cross sections (in pb) for the ``1 jet'' and ``2 jet'' sample, as
  defined in the text, computed at fixed next-to-leading order with
  \MCFM~(second column), at the LHE by the \POWHEGBOX~(third column), LHE +
  shower effects~(fourth column), \MCFM{} corrected for hadronization
  effects~(fifth column) and LHE + shower and hadronization~(last
  column). Scale-variation bands are also shown.}
\label{tab:MCFM_POWHEG} 
\end{table}
In tab.~\ref{tab:MCFM_POWHEG} we compare the \MCFM{} results computed by the
ATLAS Collaboration with our results at different stages.  The \MCFM{} NLO
results are as obtained by simply running the \MCFM{} code, i.e.~we have
subtracted all the DPI corrections and undone the hadronization corrections
applied by ATLAS.   
In this way, we can compare the results for the \MCFM{} NLO $W$+1$b$-jet and
$W$+1$b$+1$j$ production with the LHE and LHE + shower ones.  We observe a
good level of agreement among these predictions, within the scale-variation
bands. In addition, we notice that hadronization effects (last column of the
table) have a very small impact on the showered results, at the level of a
few percent.

The theoretical result for ``1+2~jet'' production with hadronization effects
included (i.e.~the sum of the numbers in the MCFM+HAD column), plus DPI
corrections, quoted in ref.~\cite{Aad:2013vka}, is given by
\begin{equation}
\sigma({\rm \mbox{1+2~jet}}) = 4.70\,^{+0.82}_{-0.65}~{\rm pb} \,.
\end{equation}
This value for the cross section is in agreement, within the scale-variation
band, with our 4-flavour result obtained with the \Wbbj+\MINLO{} generator of
eq.~(\ref{eq:sig_ATLAS_DPI}).
\begin{figure}[htb]
\begin{center}
\includegraphics[width=0.5\textwidth]{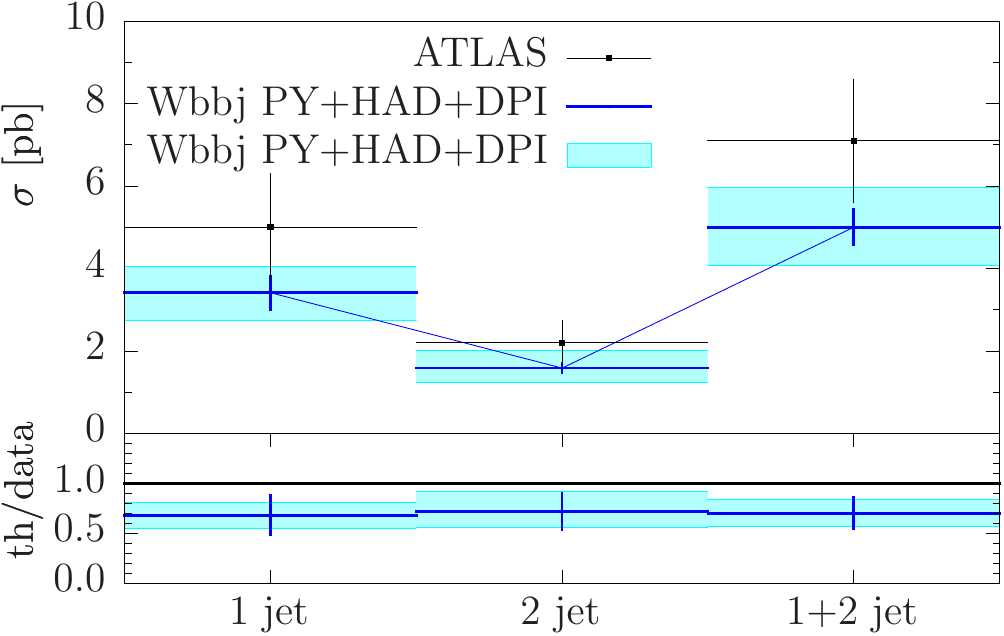}
\end{center}
\caption{Cross sections, within the cuts described in the text, for the ATLAS
  measurement of ``1 jet'', ``2 jet'' and ``1+2 jet'' $Wb$ production, using
  the scale of eq.~(\ref{eq:ATLAS_scale}) as central scale of the primary
  process.  The theoretical central results and error bands have been
  corrected for the DPI effects of tab.~\ref{tab:DPI}. Results are at the
  full shower+hadron level.}
\label{fig:WBB-WBBJ_band-ATLAS-HAD_XS-Wbj-ATLAS2-atlas2}
\end{figure}
If instead we use the scale in eq.~(\ref{eq:ATLAS_scale}) as central scale for the
primary process, we get
\begin{equation}
\sigma({\rm \mbox{1+2~jet}}) = 5.00\,^{+0.97}_{-0.93}~{\rm pb} \,,
\end{equation}
with all the other results for the ``1 jet'' and ``2 jet'' sample collected
in fig.~\ref{fig:WBB-WBBJ_band-ATLAS-HAD_XS-Wbj-ATLAS2-atlas2}. In this
figure, the cross sections obtained using eq.~(\ref{eq:ATLAS_scale}) as scale
for the primary process are displayed in blue, with the associated
statistical and DPI error bars. The bands are obtained exactly as in the
previous section, by varying the scales around the central one. With this
scale choice, we have a higher degree of overlapping of the variation bands
with the data, even if the agreement is still not perfect.

\begin{figure}[htb]
\begin{center}
\includegraphics[width=0.49\textwidth]{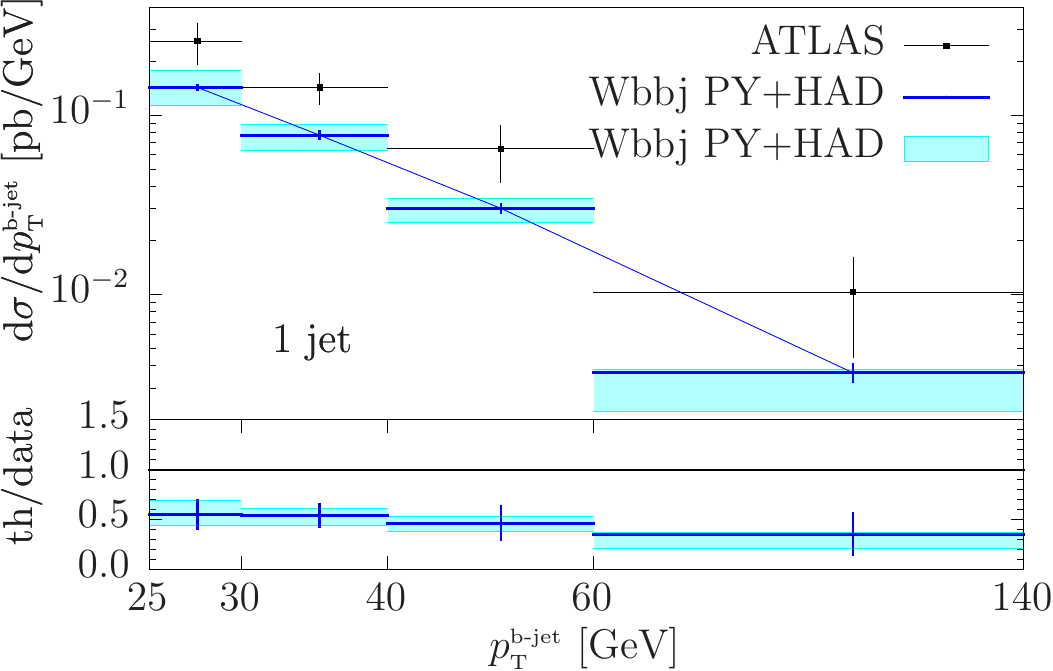}
\includegraphics[width=0.49\textwidth]{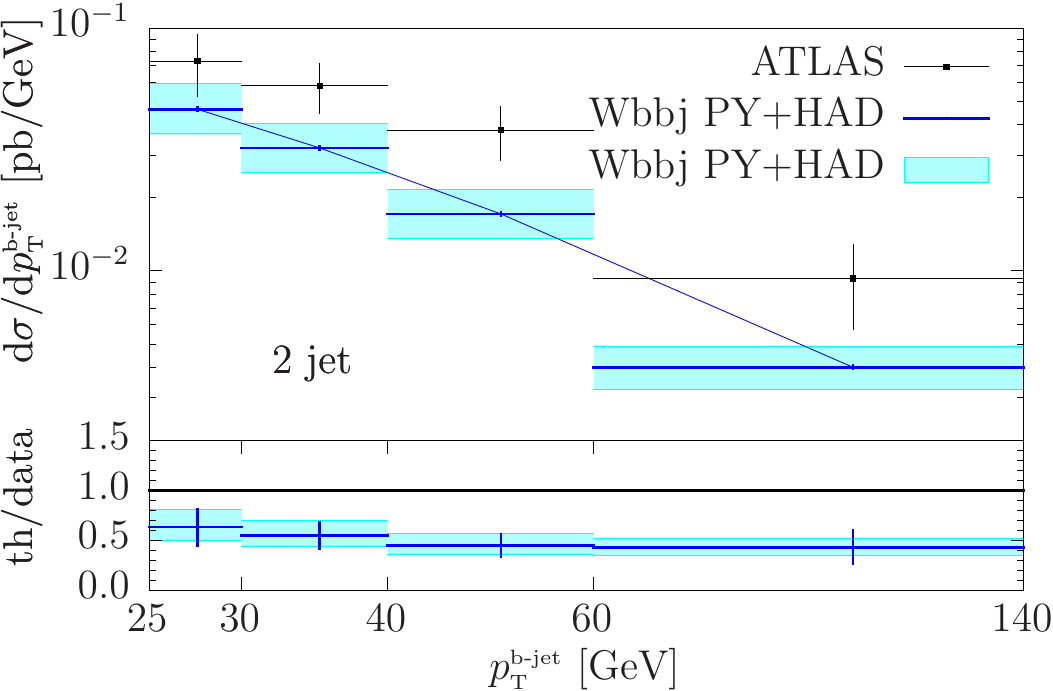}
\end{center}
\caption{Transverse-momentum distribution of the $b$-tagged jet in the ``1
  jet'' and ``2 jet'' samples, using the scale of eq.~(\ref{eq:ATLAS_scale})
  as central scale of the primary process. The theoretical results are at the
  shower+hadron level. No DPI corrections included.}
\label{fig:WBBJ_band-ATLAS-HAD_b-pt-2j-atlas2}
\end{figure}

In fig.~\ref{fig:WBBJ_band-ATLAS-HAD_b-pt-2j-atlas2}, we plot the $\pt$
spectrum of the $b$-tagged jet in the ``1 jet'' and ``2 jet'' samples. The
two panels of the figure are equivalent to
figs.~\ref{fig:WBBJ_band-ATLAS-HAD_b-pt-1j}
and~\ref{fig:WBBJ_band-ATLAS-HAD_b-pt-2j}, right panels, but with the scales
of eq.~(\ref{eq:ATLAS_scale}). Although the blue curves are 10-20\% higher
than the green ones in figs.~\ref{fig:WBBJ_band-ATLAS-HAD_b-pt-1j}
and~\ref{fig:WBBJ_band-ATLAS-HAD_b-pt-2j}, the ratio of the theoretical
predictions over the data is around the 50\% level. We recall again that DPI
corrections have not been included in these figures, and this partially
accounts for the discrepancy between theoretical results and data.

\section{Conclusions}
\label{ref:conclusions}

The production of a $W$ boson in association with one or more $b$ jets is a
significant background for $HW$ production, with the Higgs boson decaying
into $b$ quarks, and to single-top and top-pair production in the Standard
Model and to many new-physics searches.

In this paper we have presented a NLO+parton-shower event generator for
$Wb\bar{b}j$ production, where bottom-quark mass effects and spin
correlations of the leptonic decay products of the $W$ boson have been fully
taken into account.  The code has been automatically generated using the two
available interfaces of \MG{} and \GOSAM{} to the \POWHEGBOX{}.

We have applied the \MINLO{} procedure to this process and compared several
relevant distributions with the corresponding ones generated with the
NLO+parton-shower code for $Wb\bar{b}$ production. We have investigated in
detail the kinematic region where the transverse momentum of the hardest
jet $j$ in $Wb\bar{b}j$ production becomes small, and we have found good
agreement between the two codes.

We have shown results using a dynamical scale both for the \Wbb{} and
\Wbbj{} generators, and studied their factorization- and
renormalization-scale dependence.
We have compared our results with all the experimental data collected at the
LHC for $Wb(b)$ production, published in the literature up to now.  While we
found a very good agreement with the $Wb\bar{b}$ cross section measured by
the CMS Collaboration, the agreement with the ATLAS data for $Wb$ production
is less satisfactory, for both our scale choices.  We point out that the
errors on the measurements are still quite large, and more precise results
expected from the runs at higher energies will be of help in understanding
the quality of the theoretical predictions.

\vspace{.5cm}

The code for $Wb\bar{b}j$ production is available in the \POWHEGBOX{} {\tt
  V2} package, under the {\tt User-Processes-V2/Wbbj/} directory, while the
code for $Wb\bar{b}$ production is available in the {\tt
  User-Processes-V2/Wbb\_dec/} folder.  All the information can be found at
the \POWHEGBOX{} web page {\tt http://powhegbox.mib.infn.it}.

\section*{Acknowledgments}
\label{sec:Acknowledgments}
The work of G.L. was supported by the Alexander von Humboldt Foundation, in
the framework of the Sofja Kovaleskaja Award Project ``Advanced Mathematical
Methods for Particle Physics'', endowed by the German Federal Ministry of
Education and Research.  FT gratefully acknowledges support by the Italian
Ministry of University and Research under the PRIN project 2010YJ2NYW and by
the Istituto Nazionale di Fisica Nucleare~(INFN) through the Iniziativa
Specifica PhenoLNF.

We acknowledge the Rechenzentrum Garching for the computing resources used
for the calculations shown in this paper.

\appendix
\section{The change of the renormalization scheme}
\label{sec:change_scheme}
Our calculation was carried out in the mixed renormalization scheme (also
called decoupling scheme) of ref.~\cite{Collins:1978wz}, in which the light
flavours $\nlf$ (4 in our process) are subtracted in the \MSB\ scheme, while
the heavy-flavour loops are subtracted at zero momentum.  In this scheme the
heavy flavour decouples at low energy.  In fact, convergent heavy-flavour
loops are suppressed by powers of the mass of the heavy flavour. The only
unsuppressed contributions come from divergent graphs. But those are
subtracted at zero external momenta, so their contribution is removed by
renormalization for small momenta (see also ref.~\cite{Appelquist:1974tg}).

\subsection{The strong coupling constant}
In order to make contact between the renormalization carried out in the
decoupling and in the $\MSB$ scheme, we need to express the strong coupling
constant $\asnlf(\mur)$ in the mixed scheme at the scale $\mur$, with $\nlf$
light flavours, in terms of $\asnf(\mur)$ of the $\MSB$ scheme, with $\nf$
light flavours.

In the mixed scheme, charge renormalization is performed at one loop with the substitution
\begin{equation}
\label{eq:mixed}
\as^b =\mu^{2\ep} \, \asnlf(\mur) \left\{ 1 - \asnlf (\mur) \, \frac{1}{\bar\ep}
\left[ b_0^{(\nlf)} - \left(\frac{\mur^2}{m^2}\right)^\ep \frac{\TF}{3\pi}
  \right] + \ord{\as^2}\right\},
\end{equation}
where $\as^b$ is the bare coupling constant, 
\begin{equation}
b_0^{(n)}=\frac{11 \CA - 4 \, \TF \,n}{12\pi}\,,
\end{equation}
and
\begin{equation}
\frac{1}{\bar\ep} = \frac{1}{\ep}+\log(4\pi)-\gamma_E\,,
\end{equation}
where dimensional regularization in $d=4-2\ep$ has been used.
In the \MSB\ scheme the renormalization prescription is
\begin{equation}
\label{eq:MSbar}
\as^b = \mur^{2\ep} \, \asnf(\mur) \left\{ 1 - \asnf(\mur) \, \frac{1}{\bar\ep}\,
b_0^{(\nf)}+ \ord{\as^2}  \right\},
\end{equation}
where $\nf = \nlf + 1$.  

It is straightforward to check that the couplings in the two schemes obey the
well-known $\nlf$-flavour and $\nf$-flavour renormalization-group equation
respectively
\begin{eqnarray}
\label{eq:RG1}
\frac{\mathd \asnlf(\mur)}{\mathd\log\mur^2} &=& -b_0^{(\nlf)}\(\asnlf(\mur)\)^2 + \ord{\as^3},
\\
\label{eq:RG2}
\frac{\mathd \asnf(\mur)}{\mathd \log\mur^2} &=& -b_0^{(\nf)} \( \asnf(\mur)\)^2 + \ord{\as^3},
\end{eqnarray}
that can be easily derived by imposing the independence of the bare coupling
$\as^b$ under a renormalization-group transformation.
Up to order $\as^3$, the solutions of eqs.~(\ref{eq:RG1}) and~(\ref{eq:RG2})
are given by
\begin{eqnarray}
\label{eq:asnlf}
\asnlf(\mur) &=& \asnlf(m) -b_0^{(\nlf)} \as^2 \log\frac{\mur^2}{m^2}\,,
\\
\label{eq:asnf}
\asnf(\mur) &=& \asnf(m) -b_0^{(\nf)} \as^2 \log\frac{\mur^2}{m^2}\,.
\end{eqnarray}
 Combining eqs.~(\ref{eq:mixed}) and~(\ref{eq:MSbar}) we have
\begin{equation}
\label{eq:as_mixed_MSbar}
\asnlf(\mur) = \asnf(\mur) - \frac{2}{3}\,\TF\,\frac{\as^2}{2\pi}\, \log\frac{\mur^2}{m^2}
 + \ord{\as^3}\,,
\end{equation}
which is the standard \MSB{} matching condition for flavour crossing
\begin{equation}
\label{eq:as_threshold}
\asnlf(m) = \asnf(m) + \ord{\as^3}\,.
\end{equation}

\subsection{The parton distribution functions}
Similarly to what has been done for the strong coupling constant, we need to
find the changes to apply to the pdfs, in order to change scheme.  The parton
distribution functions $f_i(x,\muf)$ for $\nf$ and $\nlf$ massless flavours
must match at the mass $m$ of the heavy quark~\cite{Collins:1986mp}, i.e.~when
$\muf=m$. More specifically, in the \MSB\ subtraction
scheme, they must satisfy the following (see ref.~\cite{Cacciari:1998it} for
more details)
\begin{eqnarray}
\label{eq:pdfmatch1}
f^{(\nf)}_i(x,m) &=& f^{(\nlf)}_i(x,m)\,, \quad {\rm for\ }  i\ne h \,,\\
\label{eq:pdfmatch2}
f^{(\nf)}_h(x,m)&=&0\,, \\ 
\label{eq:pdfmatch3}
f^{(\nf)}_{\bar{h}}(x,m)&=&0\,,
\end{eqnarray}
where $h$ stands for the heavy flavour.  Using the Altarelli--Parisi
evolution equations together with the matching conditions given in
eqs.~(\ref{eq:pdfmatch1})-(\ref{eq:pdfmatch3}), one can easily find the
appropriate relations between the parton densities with $\nlf$ and $\nf$
active flavours for $\mu$ of the order of $m$.

In fact, the  Altarelli--Parisi equations for the parton densities with
$\nf=\nlf+1$ flavours are given by
\begin{equation}
\frac{\partial f^{(\nf)}_i(x,\mu) }{\partial \log \mu^2}=
\frac{\as^{(\nf)}(\mu)}{2\pi}\sum_j\int^1_x \frac{\mathd z}{z} \, f^{(\nf)}_j\(\frac{x}{z},\mu\)
P^{(\nf)}_{ij}(z)\,.
\end{equation}
Integrating both sides of this equation between $m$ and $\muf$, neglecting
terms of order $\as^2$, and using eq.~(\ref{eq:asnf}), we get
\begin{equation}
f^{(\nf)}_i(x,\muf)- f^{(\nf)}_i(x,m)=\frac{\as^{(\nf)}(m)}{2\pi}
\log\frac{\muf^2}{m^2} \sum_{j}\int^1_x \frac{\mathd z}{z}
f^{(\nf)}_j\(\frac{x}{z},m\) P^{(\nf)}_{ij}(z)\,.
\end{equation}
Since the heavy-quark parton distribution functions vanish at $\muf=m$ (see
eqs.~(\ref{eq:pdfmatch2}) and~(\ref{eq:pdfmatch3})), we can exclude them, by
putting $j\neq h,\bar{h}$ in the sum, and we can write
\begin{equation}
\label{eq:pdfpartons1}
f^{(\nf)}_i(x,\muf)= f^{(\nf)}_i(x,m)+ \frac{\as^{(\nf)}(m)}{2\pi}
\log\frac{\muf^2}{m^2} \sum_{j\neq h,\bar{h}}\int^1_x \frac{\mathd z}{z}
f^{(\nf)}_j\(\frac{x}{z},m\) P^{(\nf)}_{ij}(z)\,.
\end{equation}
For $i=h$ (or $i=\bar{h}$), eqs.~(\ref{eq:pdfmatch2}),
(\ref{eq:pdfmatch3}) and~(\ref{eq:pdfpartons1}) yield
\begin{equation}
\label{eq:pdfQ}
f^{(\nf)}_h(x,\muf) = \frac{\as^{(\nf)}(m)}{2\pi} \log\frac{\muf^2}{m^2}
\sum_{j\neq h,\bar{h}}\int^1_x \frac{\mathd z}{z}
f^{(\nf)}_j\(\frac{x}{z},m\) P^{(\nf)}_{hj}(z)\,,
\end{equation}
which shows that the heavy-quark pdf is of order $\as$.  Since an equation
similar to eq.~(\ref{eq:pdfpartons1}) holds for $\nlf$ flavours
\begin{equation}
\label{eq:pdfpartons0}
f^{(\nlf)}_i(x,\muf)= f^{(\nlf)}_i(x,m)+ \frac{\as^{(\nlf)}(m)}{2\pi}
\log\frac{\muf^2}{m^2} \sum_{j\neq h,\bar{h}}\int^1_x \frac{\mathd z}{z}
f^{(\nlf)}_j\(\frac{x}{z},m\) P^{(\nlf)}_{ij}(z)\,,
\end{equation}
we can take the difference between eqs.~(\ref{eq:pdfpartons1})
and~(\ref{eq:pdfpartons0}), and using eqs.~(\ref{eq:pdfmatch1})
and~(\ref{eq:as_threshold}), we obtain, for ${i\neq h,\bar{h}}$
\begin{eqnarray}
\label{eq:pdfgdiff}
&& f^{(\nf)}_i(x,\muf)-  f^{(\nlf)}_i(x,\muf)=
\nonumber\\
&& \qquad \qquad 
 =\frac{\as^{(\nlf)}(m)}{2\pi}
\log\frac{\muf^2}{m^2} \sum_{j\neq h,\bar{h}}\int^1_x \frac{\mathd z}{z}
f^{(\nlf)}_j\(\frac{x}{z},m\) \lq P^{(\nf)}_{ij}(z) - P^{(\nlf)}_{ij}(z)\rq\!.\phantom{aaaa}
\end{eqnarray}
The only splitting function that depends explicitly upon the number
of light flavours is the gluon splitting function, so that
\begin{equation}
P^{(\nf)}_{gg}(z) - P^{(\nlf)}_{gg}(z)= -\frac{2}{3}\,\TF\,\delta(1-z)\,,
\end{equation}
which applied to eq.~(\ref{eq:pdfgdiff}) gives
\begin{equation}
f^{(\nf)}_g(x,\muf)- f^{(\nlf)}_g(x,\muf)= -\frac{2}{3}\,\TF\,
\frac{\as^{(\nlf)}(m)}{2\pi} \log\frac{\muf^2}{m^2}\, f^{(\nlf)}_g(x,m)\,.
\end{equation}
The final results are then
\begin{eqnarray}
\label{eq:pdfh}
f^{(\nf)}_{h\, (\bar h)}(x,\muf) &=& \ord{\as} \\
\label{eq:pdfq}
f^{(\nf)}_j(x,\muf) &=& f^{(\nlf)}_j(x,\muf) +\ord{\as^2}\,, \quad {\rm for\ }
j\ne h, \bar{h}, g \\ 
\label{eq:pdfg}
f^{(\nf)}_g(x,\muf) &=& f^{(\nlf)}_g(x,\muf) \lq 1-\frac{2}{3}\TF
  \frac{\as}{2\pi} \log\frac{\muf^2}{m^2}\rq + \ord{\as^2}.
\end{eqnarray}

\subsection{Summarizing}
We have now all the ingredients to translate our formulae for $Wb\bar{b}j$
production computed in the decoupling scheme to the standard \MSB{} scheme.
At the Born level, we have contributions coming from two-quark initial states
and a quark-gluon initial state. We can then schematically write the
contributions to the Born cross section as
\begin{eqnarray}
\sigma_{qq}^{\sss\rm B} &=& f^{(\nlf)}_q(x_1,\muf) \, f^{(\nlf)}_q(x_2,\muf)
      \lq \as^{(\nlf)}(\mur)\rq^3 {\cal B}_{qq}\,, \\
\sigma_{qg}^{\sss\rm B} &=& f^{(\nlf)}_q(x_1,\muf)\, f^{(\nlf)}_g(x_2,\muf)
      \lq \as^{(\nlf)}(\mur)\rq^3 {\cal B}_{qg} \,,
\end{eqnarray}
where ${\cal B}_{qq}$ and ${\cal B}_{qg}$ are the squared Born amplitude for
the $qq$ and $qg$ initial-state channels, respectively, with three powers of
the strong coupling constant stripped off and put in front.  By using
eqs.~(\ref {eq:as_mixed_MSbar}), (\ref{eq:pdfq}) and~(\ref{eq:pdfg}) we can
write, up to order $\as^4$
\begin{eqnarray}
\sigma_{qq}^{\sss\rm B} &=& f^{(\nf)}_q(x_1,\muf) \, f^{(\nf)}_q(x_2,\muf) \lq
\as^{(\nf)}(\mur)\rq^3 
\lq 1 - 2\,\TF \frac{\as}{2\pi}\, \log \frac{\mur^2}{m^2}\rq 
{\cal B}_{qq}\,, \\
\sigma_{qg}^{\sss\rm B} &=& f^{(\nf)}_q(x_1,\muf)\, f^{(\nf)}_g(x_2,\muf)
      \lq \as^{(\nf)}(\mur)\rq^3
\lq 1+\frac{2}{3}\TF  \frac{\as}{2\pi} \log\frac{\muf^2}{m^2}\rq 
\nonumber\\
&& 
 \times \lq 1 - 2\,\TF \frac{\as}{2\pi}\, \log\frac{\mur^2}{m^2}\rq  {\cal B}_{qg} 
\nonumber\\
 &=& f^{(\nf)}_q(x_1,\muf)\, f^{(\nf)}_g(x_2,\muf)
      \lq \as^{(\nf)}(\mur)\rq^3
\nonumber\\
&& \times
\lq 1+ 2\,\TF  \frac{\as}{2\pi}\( \frac{1}{3}\log\frac{\muf^2}{m^2} -\log\frac{\mur^2}{m^2}\)\rq 
 {\cal B}_{qg}\,.
\end{eqnarray}
Applying the same change of scheme to the virtual and real contributions
would give corrections of order $\as^5$ or higher, beyond the NLO accuracy of
our calculation.
In summary, in order to change scheme, we have to:
\begin{itemize}
\item add a term 
\begin{equation}
- 2\,\TF\, \frac{\as}{2\pi} \log \(\frac{\mur^2}{m^2}\) \, {\cal
  B}_{qq}
\end{equation}
to the $qq$ channel;
\item add a term 
\begin{equation}
2\,\TF\, \frac{\as}{2\pi}\lq \frac{1}{3}\log\(\frac{\muf^2}{m^2}\)
  -\log\(\frac{\mur^2}{m^2}\) \rq {\cal B}_{qg}
\end{equation} 
to the $qg$ channel.

\end{itemize}


\providecommand{\href}[2]{#2}\begingroup\raggedright\endgroup

\end{document}